\documentclass[amsmath,superscriptaddress,
amssymb,aps,prd,floatfix,twocolumn]{revtex4}
\usepackage{graphicx}
\usepackage{mathrsfs}
\usepackage[bookmarks=false]{hyperref}
\usepackage[utf8]{inputenc}
\usepackage{dcolumn}
\usepackage{bm}
\usepackage{soul}

\begin{document}

\title{Conformal symmetry breaking in holographic QCD}

\author{Luis A. H. Mamani}%
\email{luis.mamani@ufabc.edu.br}
\affiliation {Centro de Ci\^encias Naturais e Humanas, Universidade
Federal do ABC, Avenida dos Estados 5001, 09210-580 Santo Andr\'e, SP,
Brazil}

\begin{abstract}
In this paper, we investigate a simple holographic model which describes the conformal symmetry breaking at zero temperature. The model is implemented in the context of effective holographic models for QCD described by the Einstein-dilaton equations. The realization of spontaneous conformal symmetry breaking shows a massless state in the spectrum. The existence of this state is confirmed when we compute the two-point correlation function of the scalar operator in the dual field theory, where it has a pole. On the other hand, when there is an explicit conformal symmetry breaking the massless state becomes massive, which in the dual gravitational theory is related to a deformation of the Anti-de Sitter (AdS) background by a massive scalar (dilaton) field. Moreover, we show the mass dependence on the conformal dimension of the lightest state.
\end{abstract}

\maketitle

\section{Introduction}

The study of conformal symmetry (CS) and its breaking is a wide subject 
investigated in several areas of physics. Its interest goes from: 
cosmology \cite{Hinshaw:2013}, condensed matter and critical phenomena 
\cite{Stanley:1971} and particle physics \cite{Gross:1973,Politzer:1973}, 
for example. Most systems at the critical point, for example, where second 
order phase transition occurs, have conformal symmetry, and the properties 
obtained are universal, this means that the results are valid for any system 
at the critical point. As an example consider water at the critical point, where the properties of the system may be characterized by a set of critical exponents, which are obtained through experimental methods. On the other hand, consider the Ising ferromagnet model in three-dimensions, at the critical point this model has the same set of critical exponents of water. This is a remarkable result since we are considering two different systems. One of the aims of modern physics is to understand the nature behind the universality obtained in such systems. Moreover, from the phenomenological point of view, it is also interesting to investigate what happens out of the critical points, which is equivalent to CS breaking. 
This work aims to investigate the consequences of the CS breaking on the spectrum and correlation functions. From the field theory point of view, there are a few ways how the conformal symmetry breaking may be realized. First, the symmetry may be spontaneously breaking, as a consequence, at least, one massless state emerges in the spectrum, a Nambu-Goldstone boson \cite{Goldstone:1962}. Another way to realize the spontaneous symmetry breaking is through the computation of the two-point correlation function, which has a pole at $q^2=0$, where $q$ represents the momentum. Second, through explicit symmetry breaking, in this case, the Nambu-Goldstone boson becomes massive. 

On the other hand, the Anti-de Sitter/Conformal Field Theory (AdS/CFT) correspondence \cite{Maldacena:1997re} (also known as gauge/gravity duality) is a theoretical framework where we may investigate, for example, CS breaking. The gauge/gravity duality is used to investigate some aspects that we are not able to implement using usual methods in quantum field theory, for example, in the strong coupling regime. So far, it has been used to investigate a wide range of problems in physics, problems like thermal properties of strongly coupled systems and melting of particles \cite{Andreev:2007zv,BallonBayona:2007vp,Miranda:2009uw,Mamani:2013ssa,Mamani:2018uxf}, to find transport coefficients in relativistic conformal and non-conformal plasmas \cite{Baier:2007ix,Romatschke:2009kr,Grozdanov:2015kqa,Mamani:2018qzl,Diles:2019uft,Finazzo:2014cna}, condensed matter physics \cite{Herzog:2009xv,Hartnoll:2009sz,Sachdev:2010un}, entanglement entropy (see the review \cite{VanRaamsdonk:2016exw}) and lately to investigate the interior of compact objects in astrophysics \cite{Hoyos:2016zke,Annala:2017tqz}.
In this paper, we are going to use the duality to investigate the conformal symmetry breaking mapping the underlying quantum field theory into a classical gravitational theory, which should be asymptotically AdS to use the holographic dictionary \cite{Gubser:1998bc,Witten:1998qj}. We expect the results shed some new light on understanding the CS breaking in holographic models for QCD describing color confinement in the infrared (IR) region. In the context of holography, the CS breaking was previously investigated in Refs.~\cite{Hoyos:2012xc,Hoyos:2013gma}, where there are two fixed points, one in the UV and another in the IR, while in Refs.~\cite{Bajc:2012vk,Bajc:2013wha} the backreaction on the geometry was neglected, for supersymmetric theories where CS may be breaking see for instance Ref.~\cite{Argurio:2013uba}. Thus, the approach we are going to follow below is the phenomenological bottom-up at zero temperature, and the results we obtain must be valid for these kind of models, where the CFT is deformed by a relevant operator in the dual field theory. Hence, a massive dilaton field will induce a massive scalar state in the dual field theory. On the other hand, it is worth mentioning that in the top-down models at finite temperature coming from superstring theory truncation proposed by Gubser et al. \cite{DeWolfe:2010he, DeWolfe:2011ts}, conformal symmetry is preserved, and the presence of a massive dilaton does not necessarily imply explicit conformal symmetry breaking in the dual field theory. This is true even though when charge is added in the five-dimensional effective action, i.e., the black hole solution  of Einstein-Maxwell-Dilaton action, see also Refs. \cite{Finazzo:2016psx, Critelli:2017euk, Critelli:2018osu} where the authors investigated properties of the dual field theory at the critical point.

The article is organized as follows. In Section \ref{Sec:GravModel} we present the holographic model, where the five-dimensional action describes the coupling of the metric and dilaton field at zero temperature. We also explain our approach, which interpolates the dilaton field. Moreover, the background equations are solved numerically. In Section \ref{Sec:Spectrum} we calculate the spectrum of the scalar and tensor sectors, we observe for the first time the emergence of the massless mode in the scalar sector. We also determine an analytic approximation for the mass of the lightest state as a function of the conformal dimension. At the end of this Section we compute the spectrum of the tensor sector. Section \ref{Sec:CSB} contains some solutions of the gravitational background for the massless dilaton. Implications on the dual field theory are also discussed. Section \ref{Sec:Fluctuations} is devoted to show the perturbations equations on the background fields using another approach, which introduces the domain wall coordinate and the superpotential formalism. We implement the analysis in both sectors of the perturbations considered in previous Sections. At the end, we write the perturbation equations as second-order differential equations. These equations are solved using two methods, one perturbative solution and another analytic solution, which is obtained in the asymptotic region. Using a matching procedure we find a relation between the coefficients of the UV and IR solutions. In Section \ref{Sec:CorrelationFunc} we expand the on-shell action up to second order in the perturbations to obtain the two-point correlation functions, where, as expected, there is a massless pole in the scalar sector. We conclude in Section \ref{Sec:Conclusion}. Finally, complementary material are left in Appendices \ref{AppendixA} and \ref{Sec:AppendixB}.

\section{Holographic model}
\label{Sec:GravModel}
The holographic model we are going to work with is described by the gravitational action defined in Einstein frame 
\noindent
\begin{equation}\label{Eq:5DAction}
S_{{\text E}}=-M_p^3N_c^2\int dx^5\sqrt{-g}\left(R 
-\frac{4}{3}\partial^{m}\Phi\,\partial_m\Phi+V(\Phi)\right),
\end{equation}
\noindent
where $R$ is the Ricci scalar, $\Phi$ the dilaton field and $V(\Phi)$ the dilaton potential. The equations of motion obtained from this action are given by
\noindent
\begin{equation}\label{Eq:BackgroundEqs}
\begin{split}
R_{mn}-\frac{4}{3}(\partial_{m}\Phi)(\partial_{n}\Phi)
+\frac{1}{3}g_{mn}V(\Phi)&=0,\\
\frac{1}{\sqrt{-g}}\partial_m\left(\sqrt{-g}\,g^{mn}\partial_{n}\Phi\right)
+\frac{3}{8}\partial_{\Phi}V(\Phi)&=0.
\end{split}
\end{equation}
\noindent

On the other hand, in holographic QCD at zero temperature we use an ansatz for the metric, which is given by
\noindent
\begin{equation}\label{Eq:BackgroundMetric}
ds^2=e^{2A(z)}\left(dz^2+\eta_{\mu\nu}dx^{\mu}dx^{\nu}\right),
\end{equation}
\noindent
where $A(z)$ is the warp factor. The form of the metric guarantees Poincar\'e invariance in the transverse direction to the holographic coordinate $z$. Introducing a new function defined by $\zeta(z)\equiv e^{-A(z)}$, the Einstein and Klein-Gordon equations take the simple form \cite{Ballon-Bayona:2017sxa}
\noindent
\begin{equation}\label{Eq:Background}
\zeta''-\frac{4}{9}\,\Phi'^{\,2}\,\zeta=0,
\end{equation}
\begin{equation}\label{EqKleinGordon}
12\,\zeta'^{\,2}-3\,\zeta\,\zeta''=V,
\end{equation}
\noindent
where $'$ stands for $d/dz$. We propose to fix the profile of the dilaton field, so that by solving equation \eqref{Eq:Background} we know the warp factor, in turn, solving \eqref{EqKleinGordon} we know the potential. 

The profile of the dilaton field has the following asymptotic expansions 
in the UV ($z\to 0$) and IR ($z\to \infty$)
\noindent
\begin{equation}\label{Eq:AsympDilaton}
\begin{split}
\Phi(z)&=\phi_0\,z^{\epsilon}+G\,z^{4-\epsilon},\qquad z\to 0,\\
\Phi(z)&=C\,z^{\alpha},\qquad\qquad\qquad z\to \infty.
\end{split}
\end{equation}
\noindent
In principle, $\phi_0, G$ and $C$ are constants. However, we point out that the dilaton field is dual to a scalar operator $\mathcal{O}$ with dimension $4-\epsilon$, while the metric is dual to the energy-momentum tensor $T_{\mu\nu}$ of the dual field theory. Thus, through the holographic dictionary the constants $\phi_0, G$ and $C$ have physical interpretation. For example, $\phi_0$ is the source, $G$ is related to the vacuum expectation value (VEV) and $C$ a parameter associated with the color confinement scale. Additionally, $\epsilon=\Delta_-$ is related to the mass of the dilaton field through the relation $M^2_{\Phi}\ell^2=\epsilon(\epsilon-4)$ \cite{Gubser:1998bc}, where $\ell$ represents the AdS radius and $M_{\Phi}$ the mass of the dilaton. The profile of the dilaton close to the boundary \eqref{Eq:AsympDilaton} guarantees the correct asymptotic behavior, in agreement with what is expected from the scalar operator $\mathcal{O}$ coupled to the source (for a discussion see for instance \cite{Csaki:2006ji}, see also \cite{Ballon-Bayona:2017sxa}). 
On the other hand, in the deep IR region, the profile of the dilaton field guarantees confinement, because it satisfies the general criteria investigated in Ref.~\cite{Gursoy:2007er}, where the authors showed that linear behavior is guaranteed for $\alpha=2$. Our pivotal aim is to investigate the conformal symmetry breaking, then,  we do investigate two interesting cases: $\alpha=1$ and $\alpha=2$. The motivation for studying $\alpha=1$ is because the dilaton arising in string theory has linear behavior, which has been investigated in Refs.~\cite{Gubser:2008ny,Gubser:2008yx}.

There is a wide range of possibilities of interpolating between the UV and IR asymptotic behaviors \eqref{Eq:AsympDilaton}. We choose the simplest function
\noindent
\begin{equation}\label{Eq:Dilaton}
\Phi(z)=\phi_0\,z^{\epsilon}
+\frac{G\,z^{4-\epsilon}}{1+(G/C)\,z^{4-\epsilon-\alpha}}.
\end{equation}

\noindent
\begin{figure*}[!ht]
\centering
\includegraphics[width=7.3cm]{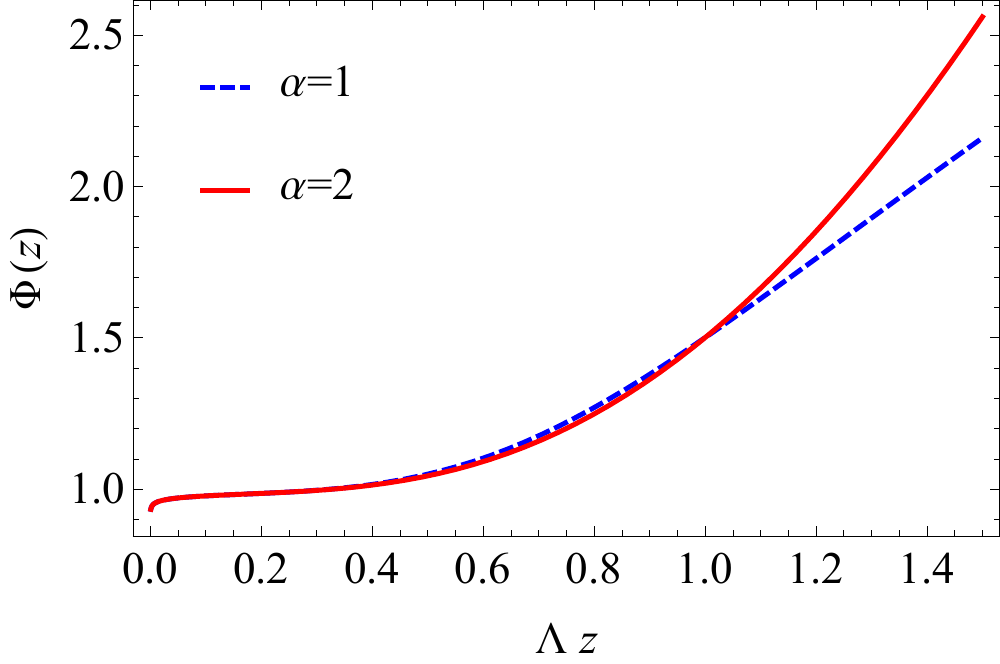}
\hspace{2cm}
\includegraphics[width=7.3cm]{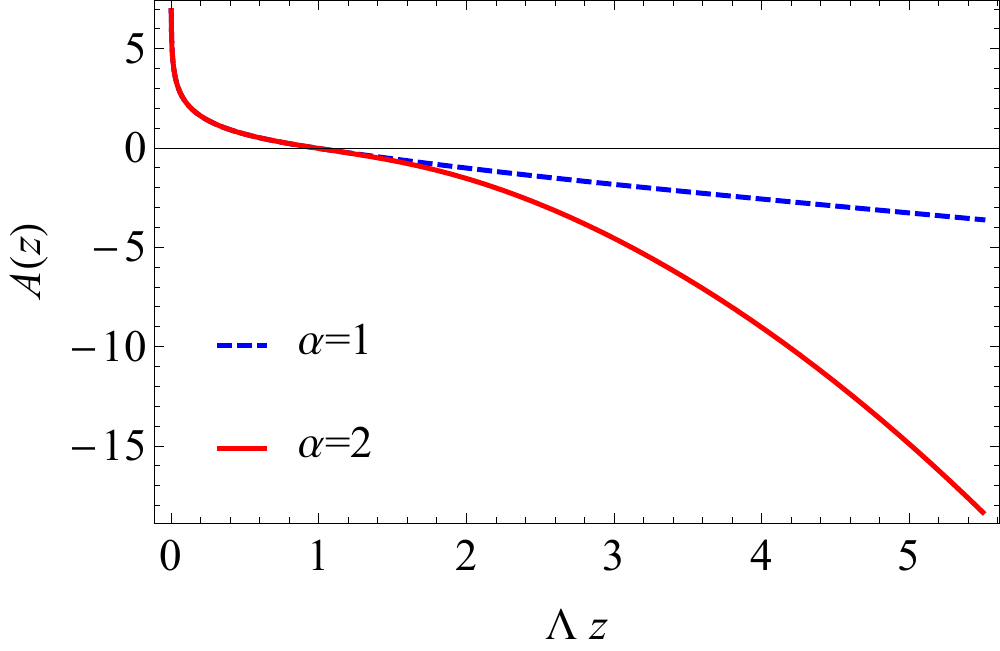}
\caption{Left: Plotting the profile of the dilaton field. 
Right: Numerical results 
of the warp factor. Both figures were 
obtained by setting: $\epsilon=0.01$, 
$\phi_0/\Lambda^{\epsilon}=1$ 
and $\ell=1$.
}\label{Fig:NumericBackground}
\end{figure*}
\noindent
\subsection{The gravitational background}
Once the dilaton field is given by \eqref{Eq:Dilaton}, the warp factor and dilaton potential are obtained by solving numerically Eqs.~\eqref{Eq:Background} and \eqref{EqKleinGordon}, respectively. The solutions determine completely the gravitational background. In the forthcoming analysis, we work with small values of $\epsilon$, i.e., $\epsilon\ll 1$. To simplify the numerical analysis, we do introduce a dimensionless coordinate $u=\Lambda z$, hence, the parameters of the model: $G$, $C$ and $\phi_0$ are normalized by the new parameter $\Lambda$. In Fig.~\ref{Fig:NumericBackground} we plot the dilaton (left panel) and warp factor (right panel). We do not use the dilaton potential, at least in the first part of the work.

\section{Spectrum}
\label{Sec:Spectrum}

\begin{figure*}[ht]
\centering
\includegraphics[width=7.2cm]{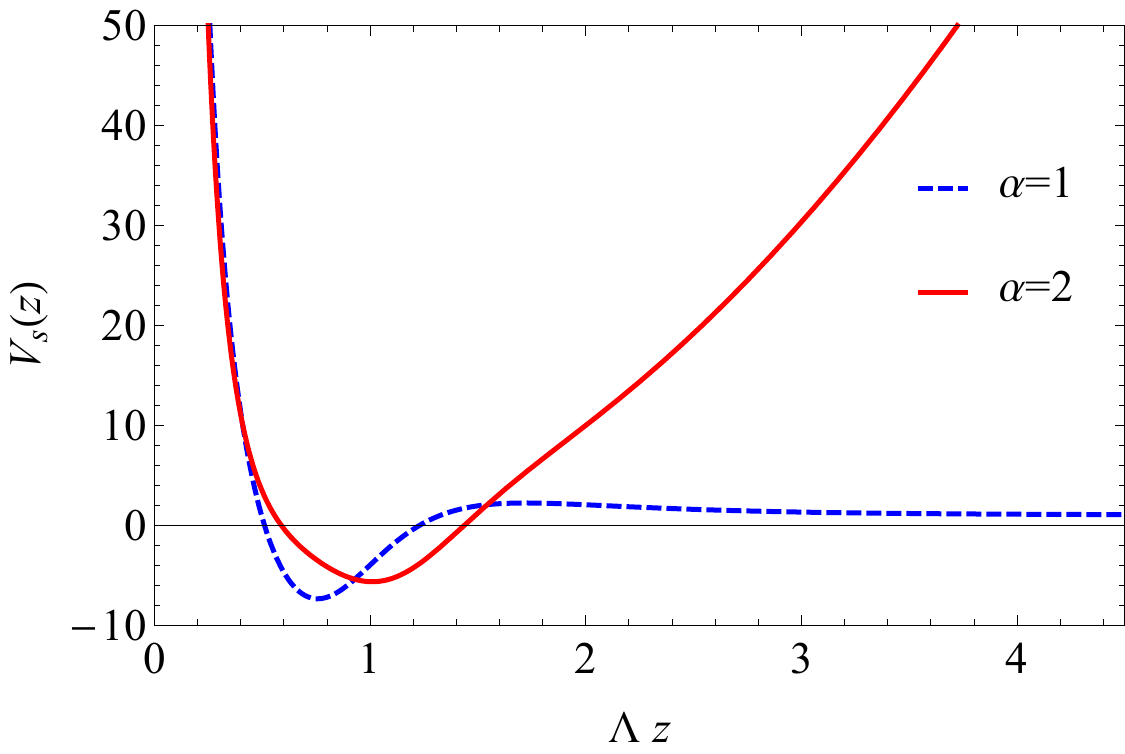}
\hspace{2cm}
\includegraphics[width=7.2cm]{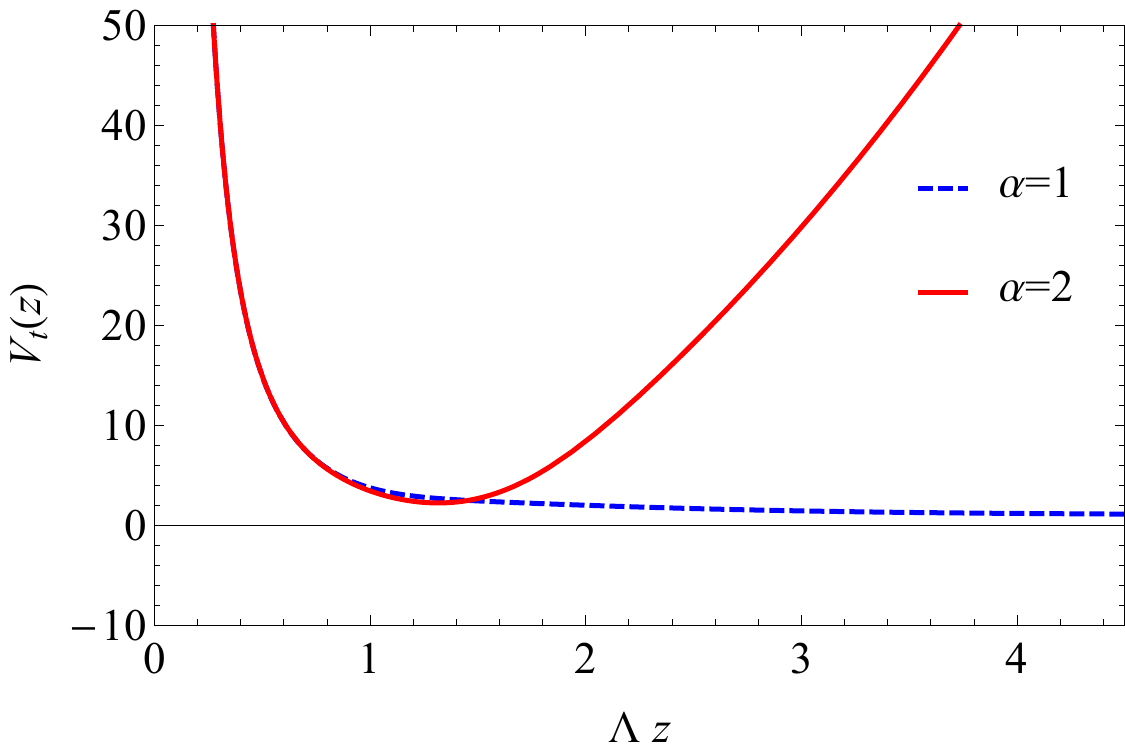}
\caption{
The figure shows the potential of the  
Schr\"odinger-like equations for the linear dilaton $\alpha=1$ 
and quadratic dilaton ($\alpha=2$).
}
\label{Fig:Mass}
\end{figure*}

The spectrum in the dual field theory is obtained from the perturbations on the metric and dilaton field around their background values, i.e., $g_{mn}\to g_{mn}(z)+h_{mn}(z,x^{\mu})$ and $\Phi\to \Phi(z)+\chi(z,x^{\mu})$. Three sectors are emerging in the perturbation equations. However, we focus on the scalar and tensor sectors because the scalar sector is associated with dual spin-zero states, while the tensor sector with spin 2 states. Moreover, in holographic QCD, the scalar sector is related to scalar glueballs due to the connection between $\mathcal{O}$ and Tr$F^2$ (the trace of the Yang-Mills gauge field), for discussion see Refs.~\cite{Gursoy:2007er,Ballon-Bayona:2017sxa}. For details on the derivation of the perturbation equations and writing them in terms of gauge-invariant variables, see for instance Refs.~\cite{Ballon-Bayona:2017sxa, Kiritsis:2006ua}. Finally, the corresponding perturbation equations in terms of gauge-invariant variables may be written as Schr\"odinger-like equations
\noindent
\begin{equation}\label{Eq:Schrodinger}
-\psi_{s,t}''(z)+V_{s,t}(z)\,\psi_{s,t}(z)=m_{s,t}^2\,\psi_{s,t}(s),
\end{equation}
\noindent
where $\psi_s(z)$ and $\psi_t(z)$ represent the wave functions of the scalar and tensor sectors, respectively. The potential of the scalar sector is given by $V_s(z)=(B_s'(z))^2+B_s''(z)$, where $2B_s(z)=3A(z)+2 \ln{|X(z)|}$ and $X(z)=\Phi'(z)/(3A'(z))$, a plot of this potential is displayed in left panel of Fig.~\ref{Fig:Mass}. On the other hand, the potential of the tensor sector is given by $V_t(z)=(B_t'(z))^2+B_t''(z)$, where $2B_t(z)=3A(z)$, a plot of the tensor potential is displayed in right panel of Fig.~\ref{Fig:Mass}.

\noindent

\subsection{Numerical solution - scalar sector}
\begin{figure}[ht]
\centering
\includegraphics[width=5.2cm]{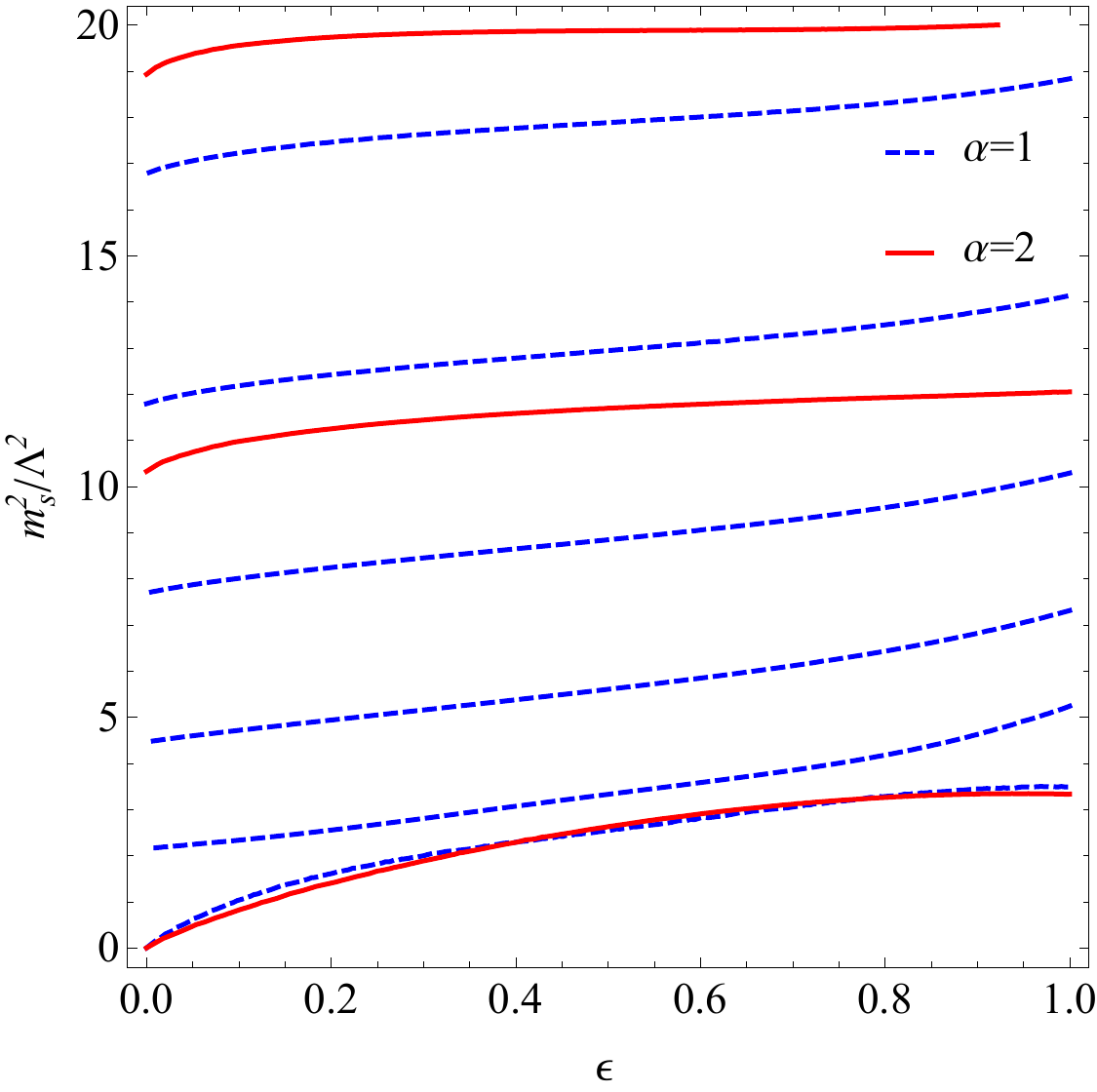}
\caption{
Figure shows the mass spectrum obtained by solving the 
Schr\"odinger-like equation using a shooting method. We 
observe that the massless state becomes lightless as the parameter 
$\epsilon$ increases.}
\label{Fig:ScalarMass}
\end{figure}

Here we obtain numerical solutions of the eigenvalue problem represented by the Schr\"odinger-like Eq.~\eqref{Eq:Schrodinger}. We solve the problem using a shooting method, the ``initial conditions'' are the asymptotic solutions close to the boundary $\psi_s\sim a_0\,z^{-\epsilon+5/2}+a_1\,z^{\epsilon-3/2}$, setting $a_1=0$ we choose the normalized solution. The results are displayed in Fig.~\ref{Fig:ScalarMass}. In this figure, we observe the emergence of a massless state in the limit $\epsilon\to 0$. On the other hand, in such a limit, the operator $\mathcal{O}$ with dimension $4-\epsilon$, becomes marginal. This result is interesting, because the massless state may be interpreted as a Nambu-Goldstone boson, consequently, a signal of spontaneous conformal symmetry breaking.

From the numerical results we realized that $m_{s}^2/\Lambda^2\sim \epsilon^{\gamma}$ for $\epsilon\ll 1$ ($\gamma$ a real number), for a fixed value of the parameter $\phi_0/\Lambda^{\epsilon}=1$. Another possibility is to fix $\phi_0/\Lambda^{\epsilon}$ using the ratio of the first glueballs states (obtained on the Lattice \cite{Meyer:2004gx}), $m_{0^{++}}=1475(30)(65)$MeV and $m_{0^{++*}}=2755(70)(120)$MeV. Then,  $\Lambda$ may be fixed by comparing the first numerical result with the corresponding first glueball state as was done in Ref.~\cite{Ballon-Bayona:2017sxa}. By doing so, the parameters are: $\phi_0/\Lambda^{\epsilon}=113.3$ and $\Lambda=420.5$MeV. However, the focus here is not to find out the spectrum and compare it with the corresponding results available in the literature, but the investigation of CS breaking, for that reason the value of these parameters is in general different.

\noindent

\subsubsection{Analytic approximation}

In a general case, the potential of the Schr\"odinger-like equation may be expanded close to its minimum located at $z=z_*$ as 
\noindent
\begin{equation}
V_s(z)=V_s(z_*)+V_s'(z_*)(z-z_*)
+\frac{V_s''(z_*)}{2}(z-z_*)^{2}+\cdots
\end{equation}
\noindent
As $z_*$ is the coordinate where the potential reaches its minimum, thus, $V_s'(z_*)$ should vanish at this point. Moreover, $V_s''(z_*)>0$ (see Fig.~\ref{Fig:Mass}), whereas $V_s(z_*)$ may be positive, negative or zero. Thus, the Schr\"odinger-like equation becomes (neglecting terms larger than $\mathcal{O}((z-z_*)^2)$)
\noindent
\begin{equation}
-\psi''_s(z)+\frac{V_s''(z_*)}{2}(z-z_*)^2\psi_s(z)
=\left(m_s^2-V_s(z_*)\right)\psi_s(z).
\end{equation}
\noindent
Defining a new variable by  
$\xi=r\sqrt[4]{V_s''(z_*)/2}$, where $r=z-z_*$, the last equation 
takes the form 
\noindent
\begin{equation}\label{EqSchroNew}
-\psi_s''(\xi)+\xi^2\psi_s(\xi)
=\lambda\,\psi_s(\xi),
\end{equation}
\noindent
where $\lambda=\left(m_s^2-V_s(z_*)\right)\sqrt{2/V_s''(z_*)}$. The normalizable 
asymptotic solution of Eq.~\eqref{EqSchroNew} is $\psi_s\sim e^{-\xi^2/2}$, then, introducing a new regular function $g(\xi)$ defined by $g(\xi)=\psi_s(\xi)\, e^{\xi^2/2}$, the new differential equation reads as 
\noindent
\begin{equation}
g''(\xi)-2\,\xi\,g'(\xi)+(\lambda-1)g(\xi)=0.
\end{equation}
\noindent
The general solution of this equation may be written as
\noindent
\begin{equation}
g(\xi)=C_1 H_{\frac{\lambda-1}{2}}(\xi )
+C_2 \, _1F_1\left(\frac{1-\lambda}{4};\frac{1}{2};\xi ^2\right)
\end{equation}
\noindent
where: $H_{\frac{\lambda-1}{2}}(\xi)$ is the Hermite polynomial and $_1F_{1}(a;b;\xi^2)$ the Kummer confluent hypergeometric function. On the other hand, the regularity condition of $\psi(\xi)$ fix the constant $C_2=0$ and $(\lambda-1)/2=n$, where $n=0,1,2\cdots$. Then, the expression for the mass is given by
\noindent
\begin{equation}\label{Eq:MassExpression}
m_s^2= V_s(z_*)+(1+2\,n)\sqrt{\frac{V_s''(z_*)}{2}}, 
\quad (n=0,1,\cdots)
\end{equation}
\noindent
where we have reestablished the original coordinate. We point out that the potential has an implicit dependence on the conformal dimension, moreover, $z_*$ is also a function of the conformal dimension and may be determined from the condition $V'(z_*)=0$. It is worth mentioning that Eq.~\eqref{Eq:MassExpression} is a general result and may be used when an analytic expression of the potential is known. For example, applying \eqref{Eq:MassExpression} for the holographic model describing scalar glueballs \cite{Colangelo:2009ra} we obtain the expression $m_s^2=4+\sqrt{15}+4n$, which is a good approximation of the exact solution $m_s^2=8+4n$. Another example is given by the spectrum of the scalar mesons \cite{Vega:2008af}, $m_s^2=4+\sqrt{3}+4n$, which is a good approximation of the exact solution  $m_s^2=6+4n$.

\begin{figure}[ht]
\centering
\includegraphics[width=5.2cm]{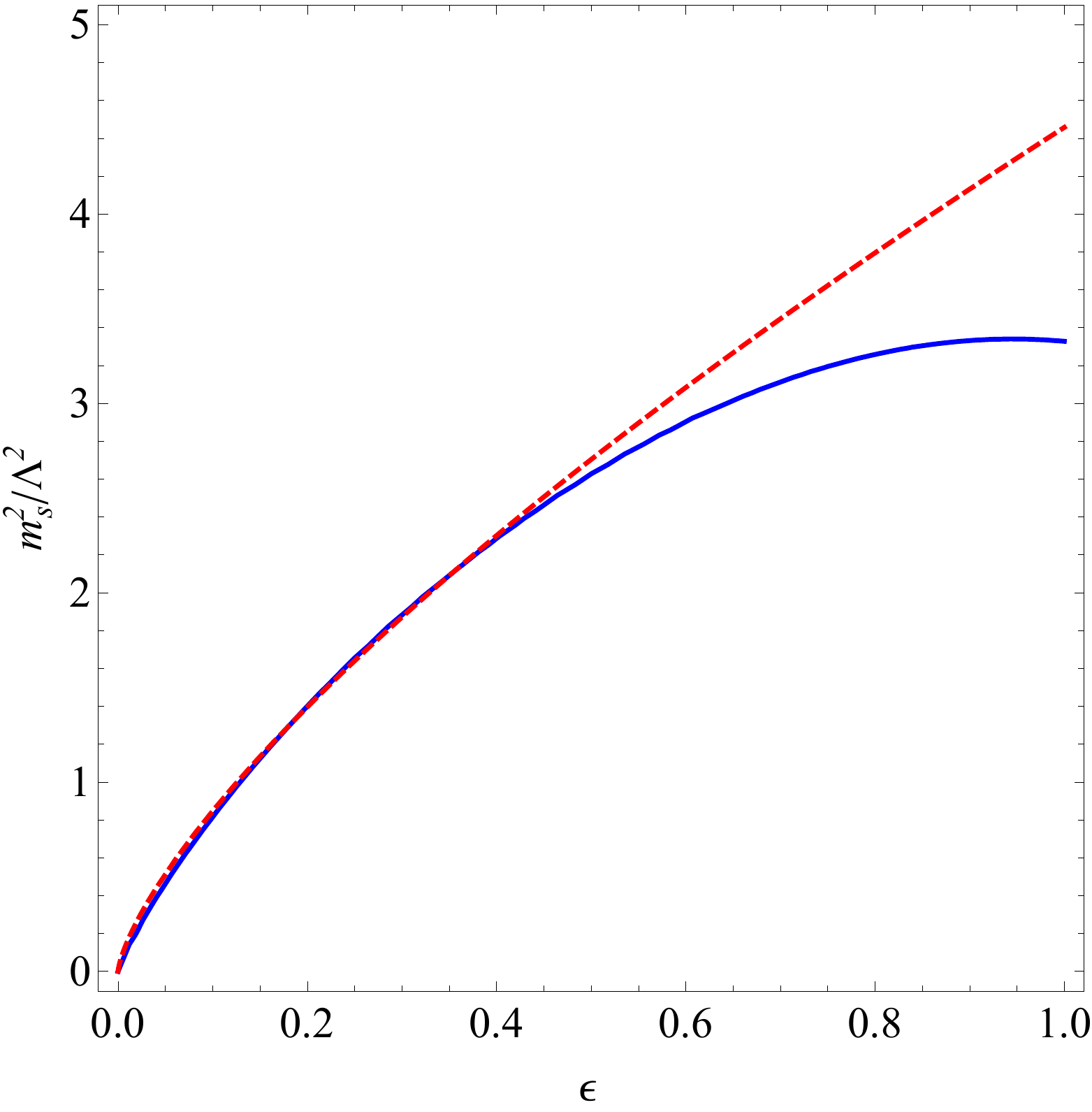}
\caption{Figure showing the behavior of the mass  
as a function of the conformal dimension for $\phi_0/\Lambda^{\epsilon}=1$.}
\label{Fig:Cond}
\end{figure}

However, in our case the potential is obtained numerically, hence, Eq.~\eqref{Eq:MassExpression} may be used to obtain an approximation for the mass as a function of the conformal dimension, $\epsilon$. In the following analysis we set $\phi_0/\Lambda^{\epsilon}=1$ and $\alpha=2$, so that we obtain $V_s(z_*)$ for $\epsilon\in [0,0.3]$, at the end we replace in Eq.~\eqref{Eq:MassExpression}, then, we fit the result to get an analytic approximation for $m_s^2(\epsilon)$, which takes the form 
\noindent
\begin{equation}\label{Eq:MassApprox1}
m_s^2= 5\,\epsilon^{4/5}, \qquad 0\leq\epsilon\leq 0.3,
\end{equation}
\noindent
An overlap of the numerical and analytic result, Eq.~\eqref{Eq:MassApprox1}, is shown in Fig.~\ref{Fig:Cond}, where the dashed line shows the analytic solution, while the continuous line shows the numerical solution. We observe a good agreement in the region of interest.

\noindent

\subsection{Numerical solution - tensor sector}
\begin{figure}[ht]
\centering
\includegraphics[width=5.2cm]{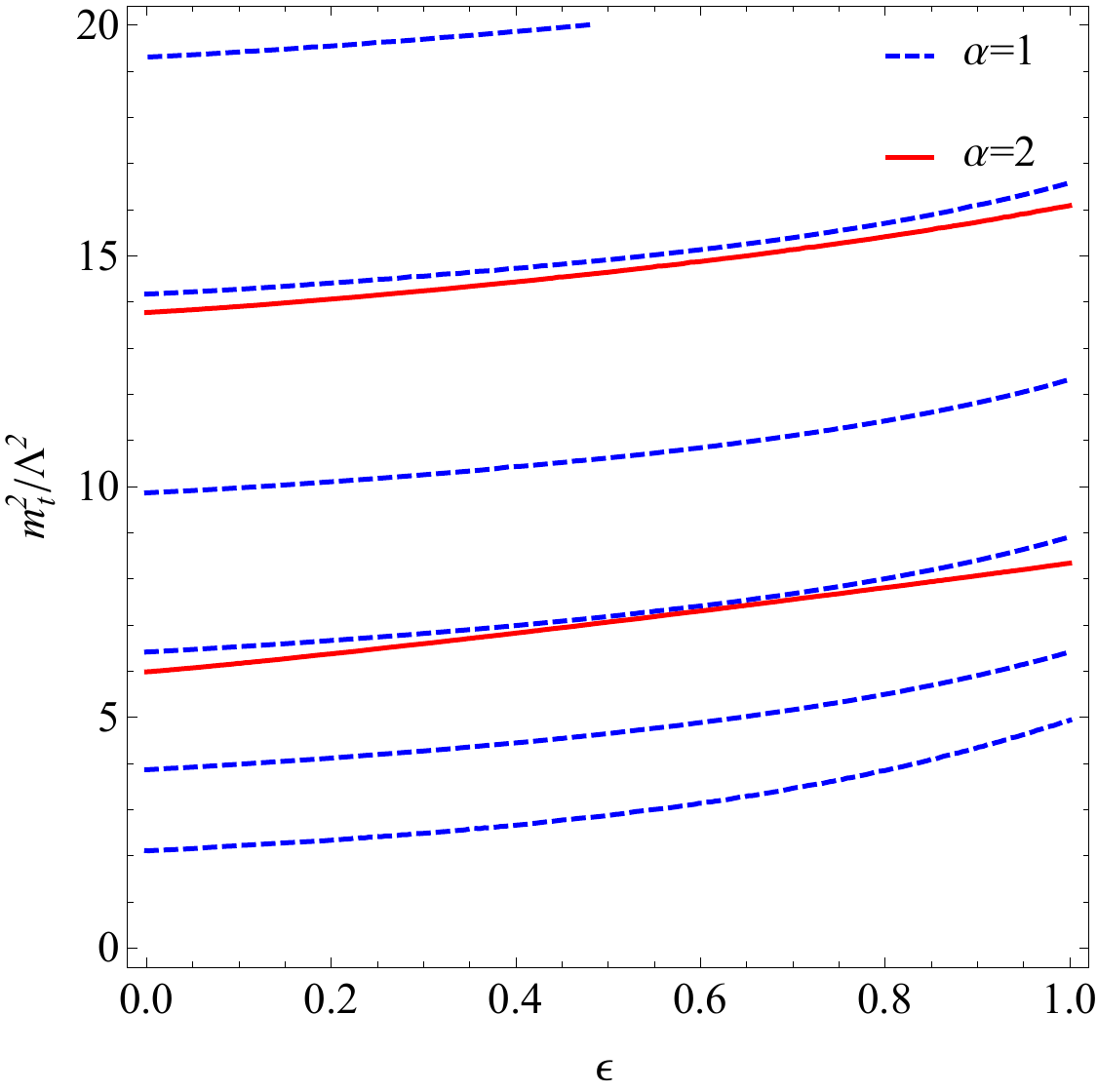}
\caption{
The figure shows the mass spectrum obtained by solving the Schr\"odinger-like equation using a shooting method. We do not observe any massless state in this sector.
}
\label{Fig:TensorMass}
\end{figure}

Analogously to the scalar sector, we may obtain numerical solutions of the eigenvalue problem represented by the Schr\"odinger-like Eq.~\eqref{Eq:Schrodinger}. We solve the problem using a shooting method, where the ``initial conditions'' are the asymptotic solutions close to the boundary $\psi_t\sim b_0\,z^{5/2}+b_1\,z^{-3/2}$. For normalizable solutions we set $b_1=0$. The results are displayed in Fig.~\ref{Fig:TensorMass}. In this figure, we observe the dependence of the mass on the conformal dimension, in the limit of $\epsilon\to 0$ we do not observe any massless mode in this sector for both, linear and quadratic dilaton.

\noindent

\section{Conformal symmetry breaking}
\label{Sec:CSB}

The picture of the five-dimensional action \eqref{Eq:5DAction} in the dual field theory was previously investigated in Refs.~\cite{Gubser:2008ny,Gubser:2008yx} (see also \cite{Ballon-Bayona:2017sxa}). In the extreme UV, which is equivalent to be at the boundary in the bulk theory, the field theory has conformal symmetry when the dimension of the operator $\mathcal O$ is $\Delta_+=4$ (marginal operator). However, introducing a new dimension of the operator, say $\Delta_{+}=4-\epsilon$, the conformal symmetry is deformed. Thus, the Lagrangian of the resulting deformed field theory may be written as 
\noindent
\begin{equation}
\mathcal{L}=\mathcal{L}_{\text{CFT}}+\phi_{0}\,\mathcal{O},
\end{equation}
\noindent
where $\phi_0$ is the source of the relevant operator. On the other hand, the dimension of the relevant operator $\Delta_{+}=4-\epsilon$, is translated in the dual gravity theory into the mass for the dilaton through the holographic dictionary by $\Delta_{+}(\Delta_{+}-4)=M^2_{\Phi}\ell^2$. Thus, a relevant scalar operator in the field theory is dual to a massive dilaton field in the dual gravity theory, and a marginal operator is dual to a massless dilaton.

Let us consider a massless dilaton in the bulk. This means that the dilaton potential does not have a quadratic term in its asymptotic expansion close to the boundary. We can show this statement by setting $\Delta_{+}=4$ (or $\epsilon=0$), hence, the asymptotic form of the dilaton field \eqref{Eq:AsympDilaton} takes the form
\noindent
\begin{equation}\label{Eq:QuarticDil}
\Phi(z)=\phi_0+G\,z^{4}.
\end{equation}
\noindent
Plugging this expression in \eqref{Eq:Background} we get an analytic solution, which is given by 
\noindent
\begin{equation}
\zeta(z)=\frac{z^{1/2}}{\ell}\left(\frac{3}{G}\right)^{1/8}
\Gamma\left(\frac{9}{8}\right)\,I_{\frac{1}{8}}\left(\frac{2}{3}G\,z^4\right),
\end{equation}
where $I_{1/8}(x)$ is the modified Bessel function of the first kind. The corresponding expression for the dilaton potential is given by 
\noindent
\begin{equation}\label{EqPotential}
\begin{split}
V(z)=&
\frac{4}{3\,\ell^2}\bigg[9\left(\,
_{0}F_{1}\left(;\frac{1}{8};\frac{G^{2}\,z^{8}}{9}\right)\right)^2-\\
&16\,G^2\,z^8\,\left(\,
_{0}F_{1}\left(;\frac{9}{8};\frac{G^{2}\,z^{8}}{9}\right)\right)^2 \bigg],
\end{split}
\end{equation}
\noindent
where $_0F_1(;a;x)$ is the confluent hypergeometric function. Expanding Eq.~\eqref{EqPotential} close to the boundary we get
\noindent
\begin{equation}
V(\Phi)=\frac{12}{\ell^2}
+\frac{512}{81\,\ell^2}(\Phi-\phi_0)^4+\cdots,
\end{equation}
\noindent
where we have used Eq.~\eqref{Eq:AsympDilaton}. As expected, there is no quadratic term on this expansion, which means a massless dilaton field. Besides, we point out that in this case, i.e., $\epsilon\to 0$, a massless state arises in the spectrum (see Fig.~\ref{Fig:ScalarMass}). Consequently, we might interpret this situation as representing some kind of spontaneous symmetry breaking (see for instance \cite{Peskin:1995ev}). However, for concluding so, we need to show that the VEV of the scalar operator is non-zero, i.e., $\langle\mathcal{O}\rangle\neq0$. It is worth mentioning that the massless state arising by considering a dilaton like in Eq.~\eqref{Eq:QuarticDil} was reported in Ref.~\cite{Csaki:2006ji}.

On the other hand, when $\epsilon\neq 0$ the asymptotic form of the dilaton remains as in \eqref{Eq:AsympDilaton}. It is not possible to obtain an analytic solution for $\zeta(z)$, thus, the asymptotic expansion of the dilaton potential is 
\noindent
\begin{equation}
V(\Phi)=\frac{12}{\ell^2}-\frac{4}{3}M_{\Phi}^2\Phi^2+\cdots,
\end{equation}
\noindent
where we have considered the leading term of the dilaton field $\Phi\sim \phi_0\,z^{\epsilon}$ and the relation $\epsilon(\epsilon-4)=M^2_{\Phi}\ell^2$. This situation represents a massive state in the spectrum (see Fig.~\ref{Fig:ScalarMass}). We might interpret as an explicit conformal symmetry breaking because the warp factor in no longer AdS but deformed AdS.

From the field theory point of view, Goldstone's theorem \cite{Goldstone:1962} states that massless bosons arise when a global symmetry is broken. The extension of this theorem in holography was previously investigated in Refs.~\cite{Bajc:2012vk,Hoyos:2012xc,Bajc:2013wha,Hoyos:2013gma}, where the dual conformal field theory has two fixed points, one in the UV and the other in the IR, thus, there is an RG-flow from one fixed point to the other. In the 
context of top-down holographic QCD, see for instance Ref.~\cite{Ben-Ami:2013lca}.

From the field theory perspective, consider a theory with non-vanishing VEV, i.e., $\langle\mathcal{O}\rangle\neq 0$, this means that the symmetries of the Lagrangian are not the symmetries of the VEV. Then, consider that the trace of the energy-momentum tensor is zero, i.e., $T^{\mu}_{\mu}=0$, which means the theory has conformal symmetry. Therefore, a massless particle arises in the spectrum because the symmetry was spontaneously broken. As we will see below, the holographic model we are working with shares these features.

In the family of effective models for holographic QCD, it is possible to compute the VEV of the scalar operator $\mathcal{O}$ with dimension $4-\epsilon$. Similarly, the vacuum energy is obtained from the regularized on-shell action. The general renormalized action is given by 
\noindent
\begin{equation}\label{Eq:RenAction}
S_{{\text ren}}=S_{{\text E}}+S_{{\text GH}}+S_{{\text CT}},
\end{equation}
\noindent
where $S_{{\text E}}$ is given by \eqref{Eq:5DAction}, $S_{{\text GH}}$ is the Gibbons-Hawking surface term and $S_{{\text CT}}$ the counterterms action, which cancel out the divergences. The Gibbons-Hawking surface term is defined by
\noindent
\begin{equation}
S_{{\text GH}}=M_p^3N_c^2\int_{\partial M}d^4x\sqrt{-g}\,2K,
\end{equation}
\noindent
where $g=det\{g_{\mu\nu}\}$ the determinant of the induced metric (we are considering the induced metric with $g_{\mu\nu}$, with $\mu=0,\cdots,3$). and $K$ the extrinsic curvature (see Appendix \ref{AppendixA} for details). The renormalized one-point function may be obtained from the on-shell action. For our convenience, the authors of Ref.~\cite{Ballon-Bayona:2017sxa} obtained the one-point correlation functions using a minimal subtraction scheme, thus, the VEV of the scalar operator reads as
\noindent
\begin{equation}\label{Eq:RenVEV}
\langle {\cal O} \rangle^{ren}=
\frac{16}{15} M_p^3N_c^2  (4 - \epsilon) G  \, .
\end{equation}
\noindent
While the vacuum energy is given by
\noindent
\begin{equation}\label{Eq:RNEnergy}
\langle T^{00} \rangle^{ren}=
-\frac{4}{15} M_p^3N_c^2\epsilon (4 - \epsilon) \phi_0 G  \, .
\end{equation}
\noindent
Both results may be combined to show the trace anomaly of 
CFTs
\noindent
\begin{equation}\label{Eq:TraceAnomaly}
\langle T_{\mu}^{\mu}\rangle^{ren}=
-\epsilon\,\phi_0
\langle \mathcal{O}\rangle^{ren}.
\end{equation}
\noindent
Now, let us take the limit of $\epsilon\to0$. The energy-momentum tensor becomes traceless, as can be seen from Eq.~\eqref{Eq:TraceAnomaly}, this means that the CS was restored. However, the VEV remains finite. In conclusion, there is a mechanism of spontaneously conformal symmetry breaking. Consequently, a massless state must emerge in the spectrum, as happens, see Fig.~\ref{Fig:ScalarMass}.

On the other hand, when $\epsilon\neq 0$, the trace anomaly \eqref{Eq:RNEnergy} holds. This means an explicit conformal symmetry breaking. As a consequence, the massless state becomes massive.

\section{Linear perturbations}
\label{Sec:Fluctuations}

\subsection{Scalar perturbations}

In this section, we show that there is a massless pole in the two-point correlation function related to the scalar operator $\langle \mathcal{O}\mathcal{O}\rangle$. We implement an analysis following Refs.~\cite{Hoyos:2012xc,Hoyos:2013gma}. First, we introduce the domain wall coordinate defined as $dr=e^{A}dz$. Using the domain wall coordinate the background metric \eqref{Eq:BackgroundMetric} takes the new form
\noindent
\begin{equation}\label{Eq:MetricDomainWall}
ds^2=dr^2+g_{\mu\nu}dx^{\mu}\,dx^{\nu}.
\end{equation}
\noindent
In the sequence, we introduce the superpotential formalism which will be useful at the time we write the perturbation equations. We may rewrite the second-order differential equation Eq.~\eqref{Eq:Background} as two first-order differential equations: 
\noindent
\begin{equation*}
\partial_{r}\Phi=\partial W, \qquad \partial_{r}A=-\frac{4}{9}W,
\end{equation*}
\noindent
thus, the potential \eqref{EqKleinGordon} becomes 
\noindent
\begin{equation}\label{Eq:Superpotential}
V=\frac{64}{27}W^2-\frac{4}{3}\left(\partial W\right)^2,
\end{equation}
\noindent
where $\partial$ represents the derivative with respect to the background scalar field $\partial/\partial\Phi$. Let us consider the perturbations on the background metric and scalar field in the form
\noindent
\begin{equation}\label{Eq:Perturbations}
\begin{split}
g_{\mu\nu}&=g_{\mu\nu}(r)+\delta g_{\mu\nu},\\
\Phi&=\Phi_0(r)+\delta\Phi,
\end{split}
\end{equation}
\noindent
where $g_{\mu\nu}=e^{2A(r)}\eta_{\mu\nu}$, $\delta g_{\mu\nu}=e^{2A(r)}h_{\mu\nu}(r,x^{\mu})$ and $\delta\Phi=\varphi(r,x^{\mu})$. The strategy is the following, we write the equations of motion in terms of the extrinsic curvature and induced metric defined in Eq.~\eqref{Eq:ExtrinsecCurv}, see Eqs.~\eqref{Eq:FluctuationsV1N1}-\eqref{Eq:FluctuationsV1N4}. Then, we decompose the metric perturbations using the projectors defined in Eq.~\eqref{Eq:Projectors}. Finally, the corresponding perturbations equations are projected to get \eqref{Eq:FluctuationsV4N2}, \eqref{Eq:hLhTEqs} and \eqref{Eq:hTTEq}, details of this analysis are written in Appendix~\ref{AppendixA}. In the forthcoming analysis, we consider just the scalar perturbations, i.e., the scalar piece of the metric and scalar field. By eliminating the scalar function $H$ from Eqs.~\eqref{Eq:FluctuationsV4N3} and \eqref{Eq:FluctuationsV4N4} we may write the perturbation of the scalar field as a third-order differential equation
\noindent
\begin{equation}\label{Eq:ThirdDiffEqScalar}
\partial^3\varphi+P\,\partial^2\varphi+Q\,\partial \varphi
+R\,\varphi=0,
\end{equation}
\noindent
where the coefficients are given by:
\begin{equation}
\begin{split}
P=&-\frac{8}{3}\frac{W}{\partial W}+2\frac{\partial^2 W}{\partial W},\\
Q=&-\frac{8}{3}-q^2\frac{e^{-2A}}{\partial W^2}
+\frac{128}{81}\frac{W^2}{\partial W^2}
+\frac{8}{9}\frac{W \partial^2 W}{\partial W^2}\\
&-\frac{\partial^2 W^2}{\partial W^2},\\
R=&q^2\frac{e^{-2A}\partial^2 W}{\partial W^3}
-\frac{128}{81}\frac{W^2 \partial^2 W}{\partial W^3}
+\frac{8}{3}\frac{\partial^2 W}{\partial W}\\
&+\frac{\partial^2 W^3}{\partial W^3}
-2\frac{\partial^2 W \partial^3 W}{\partial W^2}
-\frac{8}{9}\frac{W\partial^2 W^2}{\partial W^3}\\
&+\frac{8}{3}\frac{W \partial^3 W}{\partial W^2}
-\frac{\partial^4 W}{\partial W}.
\end{split}
\end{equation}
\noindent
Using the so called gauge mode (see Ref.~\cite{Hoyos:2012xc} for details) we may rewrite Eq.\eqref{Eq:ThirdDiffEqScalar} in the form (this also may be interpreted as a factorization of the third-order differential equation \cite{Mueck:2001cy})
\noindent
\begin{equation}\label{Eq:ThirOrderDiffEq}
\left(\partial^2+a_1\partial+a_0\right)
\left(\partial-\frac{\partial^2 W}{\partial W}\right)\varphi=0,
\end{equation}
\noindent
where the coefficients are given by:
\noindent
\begin{equation}
\begin{split}
a_1=&-\frac{8}{3}\frac{W}{\partial W}+3\frac{\partial^2 W}{\partial W},\\
a_0=&-\frac{8}{3}-q^2\frac{e^{-2A}}{\partial W^2}
+\frac{128}{81}\frac{W^2}{\partial W^2}
-\frac{16}{9}\frac{W \partial^2 W}{\partial W^2}\\
&+2\frac{\partial^3 W}{\partial W}.
\end{split}
\end{equation}
\noindent
Introducing the following transformation 
\noindent
\begin{equation}
\left(\partial-\frac{\partial^2 W}{\partial W}\right)\varphi=
\frac{W}{(\partial W)^2}e^{-4A}\mathcal{S},
\end{equation}
\noindent
the third-order differential Eq.~\eqref{Eq:ThirOrderDiffEq} reduces to a second-order one 
\noindent
\begin{equation}\label{Eq:SecondDiffEqScalar}
\partial^2\mathcal{S}
+\partial B\, \partial \mathcal{S}
-q^2\frac{e^{-2 A}}{(\partial W)^2}\mathcal{S}=0,
\end{equation}
\noindent
where $B$ is giving by
\noindent
\begin{equation}
B=2\ln{W}-\ln{\partial W}+\frac{8}{9}\frac{W}{\partial W}.
\end{equation}
\noindent
We point out that the warp factor in terms of the superpotential is given by $A=-4 W/(9\, \partial W)$. To compute the correlation functions associated with the scalar perturbations we need to know solutions of differential Eq.~\eqref{Eq:SecondDiffEqScalar} at least in the asymptotic regions, i.e., UV and IR. We may solve Eq.~\eqref{Eq:SecondDiffEqScalar} by knowing the asymptotic form of the superpotential in the IR region, for example. However, it is possible to solve this equation also perturbatively considering $q^2$ as a small parameter, which is true for the lightest state. In the forthcoming analysis, we solve this equation using two techniques. First, we solve the equation using $q^2$ as a perturbative parameter. Second, we solve the same equation in the IR and UV, in the end, we match the corresponding solutions in the asymptotic regions. 

The perturbative solution may be written as
\noindent
\begin{equation}
\mathcal{S}=\sum_{n=0}q^{2n}\mathcal{S}_{n}. \qquad q^2< 1
\end{equation}
\noindent
Up to second order in the perturbative parameter the solution of 
Eq.~\eqref{Eq:SecondDiffEqScalar} is given by
\noindent
\begin{widetext}
\begin{equation}\label{Eq:PertSol}
\begin{split}
\mathcal{S}=&c_2\left(1+q^2\int e^{-B(\Phi_3)}
\int\frac{e^{-2A(\Phi_2)+B(\Phi_2)}}{(\partial W(\Phi_2))^2}d\Phi_2d\Phi_3
+\mathcal{O}(q^4)\right)
\\+
&c_1\bigg(\int e^{-B(\Phi_1)}d\Phi_1
+q^2\int e^{-B(\Phi_3)}\int \frac{e^{-2 A(\Phi_2)+B(\Phi_2)}}{(\partial W(\Phi_2))^2}\int e^{-B(\Phi_1)}d\Phi_1 d\Phi_2 d\Phi_3+\mathcal{O}(q^4)\bigg).
\end{split}
\end{equation}
\end{widetext}
In principle, the result \eqref{Eq:PertSol} is valid in the whole region of interest, i.e., from the UV to the IR, it represents the wave function of the lightest state. Therefore, by knowing the superpotential it is possible to evaluate these integrals, fortunately for us, we do know the IR asymptotic form of it \cite{Gursoy:2007er}
\noindent
\begin{equation}\label{Eq:SuperPotIR}
W=W_{\infty}\Phi^{\frac{\alpha-1}{2\alpha}}e^{2\Phi/3},
\end{equation}
\noindent
where $W_{\infty}$ is a constant. The expressions for $B$ and warp factor are then given by 
\noindent
\begin{equation}
\begin{split}
B=&2\Phi-\frac{3(\alpha-1)}{4\alpha}\frac{1}{\Phi}
-\frac{\alpha-1}{2\alpha}\ln{\Phi}+\ln{\frac{3W_{\infty}^2}{2}},\\
A=&\frac{\alpha-1}{2\alpha}\ln{\Phi}-\frac{2}{3}\Phi.
\end{split}
\end{equation}
\noindent
In the following we consider the case for $\alpha=2$ only and leave the results for $\alpha=1$ in Appendix \ref{Sec:AppendixB}. Thus, the perturbative solution reads as 
\noindent
\begin{equation}\label{Eq:IRPertSol}
\begin{split}
\mathcal{S}=&c_2\left(1+\mathcal{O}(q^2)\right)
-\frac{c_1 e^{-2\Phi}\Phi^{1/4}}{3 W^2_{\infty}}\left(1
+\mathcal{O}(q^2)\right), \quad \alpha=2
\end{split}
\end{equation}
\noindent
where $\mathcal{O}(q^2)$ are subleading corrections.

On the other hand, let us focus on the asymptotic form of the differential Eq.~\eqref{Eq:SecondDiffEqScalar} in the IR region where it becomes 
\noindent
\begin{equation}\label{Eq:ScalarIR}
\partial^2\mathcal{S}+
\left(2-\frac{\alpha-1}{2\alpha}\frac{1}{\Phi}\right)\partial\mathcal{S}+
\frac{36\,\widetilde{q}^{\,2}\Phi^{2-\frac{2(\alpha-1)}{\alpha}}}{(\frac{3(\alpha-1)}{\alpha}+4\Phi)^2}\mathcal{S}=0,
\end{equation}
\noindent
where $\widetilde{q}=q/W_{\infty}$. Fortunately, their solutions may be written in terms of known functions:
\noindent
\begin{equation}\label{Eq:IRAsympSol2}
\begin{split}
\mathcal{S}=&e^{-2\Phi}\Phi^{5/4}\bigg(
c_3\,U\left(1-\frac{9\widetilde{q}^{\,2}}{8},\frac{9}{4},2\Phi\right)+\\
&c_4\,L^{5/4}_{\frac{9\widetilde{q}^{\,2}}{8}-1}(2\Phi)\bigg),\qquad \alpha=2
\end{split}
\end{equation}
\noindent
where $U(a,b,x)$ and $L^{a}_{n}(x)$ are the confluent hypergeometric and generalized Laguerre functions, respectively. 

To avoid logarithms in the last solution we fix the constant $c_3=0$ and set $c_4=c_0$. The leading terms of the series expansion of Eq.~\eqref{Eq:IRAsympSol2} are:
\noindent
\begin{equation}\label{Eq:IRSol}
\begin{split}
\mathcal{S}=&\frac{9}{16}\widetilde{q}^{\,2}c_0\bigg(\frac{\Gamma(\frac{5}{4})}{2^{1/4}}-
e^{-2\Phi}\Phi^{1/4}\bigg). \quad \alpha=2
\end{split}
\end{equation}
\noindent
By matching \eqref{Eq:IRPertSol} with \eqref{Eq:IRSol} we conclude 
that
\noindent
\begin{equation}
c_2=\frac{9}{16}\widetilde{q}^{\,2}c_0\frac{\Gamma\left(\frac{5}{4}\right)}{2^{1/4}}, \quad
c_1=\frac{27}{16}W_{\infty}^2\widetilde{q}^{\,2}c_0,
\end{equation}
\noindent
consequently their ratio does not depend on $q^2$
\noindent
\begin{equation}\label{Eq:c2c1IRRatio}
\frac{c_2}{c_1}=\frac{\Gamma\left(\frac{5}{4}\right)}{3W_{\infty}^2}.
\end{equation}
\noindent

Now, let us turn our attention to the UV. The asymptotic form of the superpotential close to the boundary may be written as \cite{Ballon-Bayona:2017sxa,Hoyos:2012xc}
\noindent
\begin{equation}\label{Eq:SuperpotUV}
W=\frac{9}{4\ell}+\frac{\Delta_{+}}{2\ell}\Phi^2.
\end{equation}
\noindent
In the forthcoming analysis, we set $\Delta_{+}=4$, which is the limit of spontaneous conformal symmetry breaking. The expression \eqref{Eq:SuperpotUV} may be obtained solving Eq.~\eqref{Eq:Superpotential} for the massless dilaton, i.e., the potential just with cosmological constant,
\noindent
\begin{equation}
W=\frac{9}{4\ell}\cosh{\frac{4}{3}\Phi}.
\end{equation}
\noindent
Hence, expanding the last result up to second order in $\Phi$ we recover Eq.~\eqref{Eq:SuperpotUV}. Knowing the superpotential we may find the function $B$ and evaluate the integrals in Eq.~\eqref{Eq:PertSol}. Therefore, including the gauge mode, $\varphi_g=C_{g} \partial W$, the perturbation function $\varphi$ takes the form
\noindent
\begin{equation}\label{Eq:UVDilaExp}
\varphi=\varphi_{b}+\varphi_{N}\Phi_{0}
+C_{N}\Phi_{0}^{3/2},
\end{equation}
\noindent
where
\noindent
\begin{equation}\label{Eq:UVDilaCoeff}
\begin{split}
\varphi_{N}=&\,2C_{g}, \\
\varphi_{b}=&\,c_2-\frac{2^{17/4}}{3^4}\Gamma\left(3/4\right)c_1,\\ 
\quad C_{N}=&\,\frac{128}{243}c_1,\\
\end{split}
\end{equation}
\noindent
Analogously, the asymptotic solutions for the metric components $h$ and $h_{T}$ are determined from Eqs.~\eqref{Eq:FluctuationsV4N3} and \eqref{Eq:FluctuationsV4N2}, 
respectively
\noindent
\begin{equation}\label{Eq:hTh}
\begin{split}
h_{T}=&h_{Tb}-\frac{8}{3}\varphi_{b}\Phi_0-\frac{4}{3}\varphi_{N}\Phi_{0}^2
-\frac{16}{15}C_{N}\Phi_{0}^{5/2},\\
h=&h_b-\frac{16}{9}\varphi_{N}\Phi_0^{2}-\frac85 C_N \Phi_0^{5/2}.
\end{split}
\end{equation}
\noindent
Replacing these results in Eq.~\eqref{Eq:FluctuationsV3N1} we get a relation between $\varphi_{N}$ and $C_N$ given by 
\noindent
\begin{equation}\label{Eq:PhiNCN}
\varphi_{N}=\frac{36}{q^2}C_N.
\end{equation}
\noindent
On the other hand, using the last equation in \eqref{Eq:UVDilaCoeff} we rewrite \eqref{Eq:PhiNCN} as
\noindent
\begin{equation}\label{Eq:CNC1}
\varphi_{N}=\frac{1}{q^2}\frac{512}{27}c_1
\end{equation}
\noindent
Finally, combining Eqs.~\eqref{Eq:c2c1IRRatio}, \eqref{Eq:UVDilaCoeff} and \eqref{Eq:CNC1} we get the following relation
\noindent
\begin{equation}\label{Eq:PhiNPhib}
\varphi_N=\mathcal{F}_s(q)\,\varphi_b,
\end{equation}
\noindent
where
\noindent
\begin{equation}
\mathcal{F}_s(q)=\frac{512}{\left(\frac{3^2\Gamma(5/4)}{W_{\infty}^2}
-\frac{2^{17/4}\Gamma(3/4)}{3}\right)}\frac{1}{q^2}.
\end{equation}
\noindent

Additionally, we write $h_T$ and $h_L=h-h_T$ using \eqref{Eq:hTh} 
in the form 
\noindent
\begin{equation}
\begin{split}
h_T=&\,{h_T}_b+{h_T}_N\Phi_0+\cdots,\\
h_L=&\,{h_L}_b+{h_L}_N\Phi_0+\cdots,
\end{split}
\end{equation}
\noindent
where we have defined 
\noindent
\begin{equation}\label{Eq:hNphib}
\begin{split}
{h_T}_N=&\mathcal{F}^{T}(q) \varphi_b,\\
{h_L}_N=&\mathcal{F}^{L}(q) \varphi_b
\end{split}
\end{equation}
\noindent
with $\mathcal{F}^{T}(q)=-8/3$ and $\mathcal{F}^{L}(q)=8/3$.

From the above results, it is possible to say that the two-point correlation function associated with the scalar operator $\mathcal{O}$ has a massless pole. Using an argument similar to the one used investigating quasinormal modes, see Ref.~\cite{Kovtun:2005ev}. The statement said that the two-point correlation function is proportional to the ratio of the coefficients $\mathcal{A}$ and $\mathcal{B}$ when the perturbation function, here represented by $Z$, may we written as
\noindent
\begin{equation}\label{Eq:QNMApproach}
Z(r)=\mathcal{A}(q)\varphi_{1}(r)+\mathcal{B}(q)\varphi_{2}(r),
\end{equation}
\noindent
where the coefficients may depend on the momentum and $\varphi_{1}$($\varphi_{2}$) is the nonnormalizable(normalizable) solution close to the boundary. Hence, the two-point correlation function is given by 
\noindent
\begin{equation}
\langle\mathcal{O}\mathcal{O}\rangle\propto\frac{\mathcal{B}(q)}{\mathcal{A}(q)}
\end{equation}
\noindent
Using the same argument in our case, from Eq.~\eqref{Eq:UVDilaExp}, $\mathcal{A}=\varphi_b$ and $\mathcal{B}=\varphi_{N}$. Then, replacing \eqref{Eq:c2c1IRRatio} we get 
\noindent
\begin{equation}
\frac{\varphi_{N}}{\varphi_b}=\frac{1536}
{\frac{3^3}{W_{\infty}^2}\Gamma\left(\frac{5}{4}\right)
-2^{17/4}\Gamma\left(\frac34\right)}\frac{1}{q^2}.
\end{equation}
\noindent
Therefore, 
\noindent
\begin{equation}\label{Eq:TwoPointScalarQN}
\langle\mathcal{O}\mathcal{O}\rangle\propto\frac{1}{q^2}.
\end{equation}
\noindent
This proofs the existence of the massless pole as commented previously. An analogous discussion is also presented in Ref.~\cite{Hoyos:2013gma}. However, to show the consistency of our results we are going to write the on-shell action and find the two-point correlation function through the functional derivative.

\subsection{Tensor Perturbations}
\label{Sec:TensorPerturb}

In this section, we investigate the transverse and traceless sector (or spin 2 for short). It is not difficult to show that this sector decouples from the other sectors and its equation of motion may be written as (see Appendix \ref{AppendixA} for details)
\noindent
\begin{equation}\label{Eq:TTsector}
\partial^2 h_{\mu\nu}^{TT}+
\partial B^{TT}\partial h_{\mu\nu}^{TT}-
\frac{q^2\,e^{-2A}}{(\partial W)^2}h_{\mu\nu}^{TT}=0,
\end{equation}
\noindent
where $B^{TT}=\log{\partial W}+4 A$, which is given by (we focus on the case $\alpha=2$)
\noindent
\begin{equation}\label{Eq:BTTSector}
B^{TT}=\log{\left(\frac{2W_{\infty}}{3}\right)}
+\frac54 \log{\Phi}-2 \Phi+\frac{3}{8\Phi}
\end{equation}
\noindent
Similarly to what we have done above, the perturbative solution of Eq.~\eqref{Eq:TTsector} is given by
\noindent
\begin{equation}\label{Eq:PertTTSector1}
h^{TT}_{\mu\nu}={h_b}^{TT}_{\mu\nu}
+{h_N}^{TT}_{\mu\nu}\int e^{-B^{TT}(\Phi_1)}d\Phi_1
+\mathcal{O}(q^2),
\end{equation}
\noindent
where ${h_b}^{TT}_{\mu\nu}$ is the non-normalizable solution and ${h_N}^{TT}_{\mu\nu}$ the normalizable one. Plugging \eqref{Eq:BTTSector} into \eqref{Eq:PertTTSector1} and performing the integral we obtain
\noindent
\begin{equation}\label{Eq:PertSolTTSector}
h^{TT}_{\mu\nu}={h_b}^{TT}_{\mu\nu}
+\frac{3{h_N}^{TT}_{\mu\nu}}{4W_{\infty}}e^{2\Phi}\Phi^{-5/4}
+\mathcal{O}(q^2).
\end{equation}

On the other hand, 
the differential equation \eqref{Eq:TTsector} in the IR reduces to
\begin{equation}
\partial^2 h_{\mu\nu}^{TT}+
\left(-2+\frac{5}{4\Phi}\right)\partial h_{\mu\nu}^{TT}+
\frac{9\widetilde{q}^{\,2}}{4\Phi}h_{\mu\nu}^{TT}=0,
\end{equation}
\noindent
where we have used the superpotential \eqref{Eq:SuperPotIR} and $\widetilde{q}=q/W_{\infty}$. The solution of this differential equation is given by
\noindent
\begin{equation}\label{Eq:SolTTSector}
h^{TT}_{\mu\nu}={C_1}_{\mu\nu} U\left(-\frac{9\widetilde{q}^{\,2}}{8},\frac{5}{4},2\Phi\right)
+{C_2}_{\mu\nu}L^{1/4}_{9\widetilde{q}^{\,2}/8}(2\Phi),
\end{equation}
\noindent
where $U(a,b,x)$ and $L^{a}_{n}(x)$ are the confluent hypergeometric and generalized Laguerre functions, respectively. In order to get an expression to compare with the perturbative solution \eqref{Eq:PertSolTTSector}, we expand \eqref{Eq:SolTTSector} and set $-{C_1}_{\mu\nu}={C_0}_{\mu\nu}={C_2}_{\mu\nu}$, getting 
\noindent
\begin{equation}
h^{TT}_{\mu\nu}=\frac{9{C_0}_{\mu\nu}(\gamma_{E}+i\pi)q^2}{8W_{\infty}^2}
-\frac{9{C_0}_{\mu\nu}\Gamma(5/4)q^2}{2^{9/4} W_{\infty}^2}e^{2\Phi}\Phi^{-5/4}
+..
\end{equation}
\noindent
where $\gamma_E$ is the Euler's constant. Therefore, after matching with \eqref{Eq:PertSolTTSector} a relation between the non-normalizable and normalizable coefficients is determined
\noindent
\begin{equation}\label{Eq:hbhtt}
{h_N}^{TT}_{\mu\nu}=
-\frac{2^{3/4}\,W_{\infty}\Gamma(5/4)}{3(\gamma_E+i\pi)} {h_b}^{TT}_{\mu\nu}. 
\end{equation}
\noindent
Analogously to what we have done in the scalar case, we may obtain the two-point correlation function associated with the energy-momentum tensor, which is proportional to 
\noindent
\begin{equation}
\langle{T}{T}\rangle\propto\frac{{h_N}^{TT}}{{h_b}^{TT}}
=\frac{2^{3/4}\,W_{\infty}\Gamma(5/4)}{3(\gamma_E+i\pi)}.
\end{equation}
\noindent
Therefore, the last result shows no dependence on the momentum and no massless pole. This result is in agreement with the spectrum displayed in Fig.~\ref{Fig:TensorMass}.

On the other hand, the perturbative solution close to the boundary is also determined from the integral in \eqref{Eq:PertTTSector1} and using the superpotential \eqref{Eq:SuperpotUV}
\noindent
\begin{equation}
h^{TT}_{\mu\nu}=
{h_b}^{TT}_{\mu\nu}
+{h_N}^{TT}_{\mu\nu}\frac{\Phi_0}{4}\left(1+\frac{4\Phi_0^2}{27}\right),
\end{equation}
\noindent
plugging \eqref{Eq:hbhtt} in the last equation we get
\noindent
\begin{equation}\label{Eq:hNhb}
h^{TT}_{\mu\nu}={h_b}^{TT}_{\mu\nu}\left(1
+\mathcal{F}^{TT}(q)\,\Phi_0+\cdots\right),
\end{equation}
\noindent
where 
\begin{equation}\label{Eq:FTT}
\mathcal{F}^{TT}(q)=
-\frac{W_{\infty}\Gamma(5/4)}{3\times2^{5/4}\,(\gamma_E+i\pi)}. 
\end{equation}

\section{Two-point functions}
\label{Sec:CorrelationFunc}
To find the correlations functions we need to expand the on-shell action up to second order in the perturbations. To get finite expression we need to add counterterms to cancel out the divergences arising in the UV. Thus, the counterterms action in \eqref{Eq:RenAction} may be written as
\noindent
\begin{equation}
S_{{\text CT}}=\frac{8}{3}\,M_p^3N_c^2\int d^4x\sqrt{-g}\,\widetilde{W}(\Phi),
\end{equation}
\noindent
where $\widetilde{W}$ is a function that has the same asymptotic expansion of the superpotential close to the boundary, however, their coefficients are in general different. Hence, close to the boundary, it has the asymptotic expansion
\noindent
\begin{equation}\label{Eq:Superpotential2}
\widetilde{W}=\frac94+(4-\Delta_{+})\Phi^2.
\end{equation}
\noindent
To expand the on-shell action up to second order in the perturbations we follow the analysis implemented in Refs.~\cite{Mueck:2001cy, Hoyos:2012xc,Skenderis:2002wp,Papadimitriou:2004rz}, and write the action as 
\noindent
\begin{equation}\label{Eq:OnShellAc}
\begin{split}
S[g_{\mu\nu}+&\delta g_{\mu\nu},\Phi_0+\delta\Phi]=
S[g_{\mu\nu},\Phi_0]+\\
&\delta S[g_{\mu\nu}+\frac12\delta g_{\mu\nu},\Phi_0+\frac12\delta\Phi;
\delta g_{\mu\nu},\delta\Phi],
\end{split}
\end{equation}
\noindent
where $\delta S$ is given by
\noindent
\begin{equation}
\begin{split}
&\delta S[g_{\mu\nu},\Phi_0;\delta g_{\mu\nu},\delta\Phi]=\\
&-M_p^3N_c^2\int d^4x\sqrt{-g}\bigg\{\left(K^{\mu\nu}
-Kg^{\mu\nu}-\frac{4}{3}\widetilde{W}g^{\mu\nu}\right)\delta g_{\mu\nu}\\
&+\frac83\left(\partial_{r}\Phi_0-\partial \widetilde{W}\right)\delta\Phi\bigg\}.
\end{split}
\end{equation}
\noindent
It is worth mentioning that in the analysis we are performing, there is an implicit limit $\Phi_0\to 0$ on the on-shell action.
Setting $\Delta_+=4$ we simplify the last result, while the superpotential \eqref{Eq:Superpotential2} reduces to a constant.

Therefore, the on-shell action up to second order in the fluctuations reads as 
\noindent
\begin{equation}
\begin{split}
S&=-M_p^3N_c^2\int d^4x\sqrt{-g}\bigg(
\frac{4}{3}\partial_r\varphi \varphi-\frac43\partial^2 \widetilde{W}\varphi^2
\\&-\frac23\partial \widetilde{W}\varphi h^{\mu}_{\mu}
+\frac14h_{\mu\nu}\partial_r h^{\mu\nu}
-\frac14 h^{\mu}_{\mu}\partial_r h^{\nu}_{\nu}\bigg).
\end{split}
\end{equation}
\noindent
Now we rewrite the last result in terms of the superpotential
\noindent
\begin{equation}
\begin{split}
S&=-M_p^3N_c^2\int d^4x\, e^{4A}\partial W\bigg(
\frac{4}{3}\partial \varphi \varphi
-\frac43\frac{\partial^2 \widetilde{W}}{\partial W}\varphi^2
\\&-\frac23\frac{\partial\widetilde{W}}{\partial W}\varphi h^{\mu}_{\mu}
+\frac14h_{\mu\nu}\partial h^{\mu\nu}
-\frac14 h^{\mu}_{\mu}\partial h^{\nu}_{\nu}\bigg).
\end{split}
\end{equation}
\noindent
Hence, the finite piece of the on-shell action becomes
\noindent
\begin{equation}
S=-M_p^3N_c^2\int d^4x\,\bigg(
\frac{16}{3}\varphi_N \varphi_b+{h_N}_{\mu\nu}{h_b}^{\mu\nu}
-{h_N}^{\mu}_{\mu}{h_b}^{\nu}_{\nu}\bigg)
\end{equation}
\noindent
Using the decomposition \eqref{Eq:hDecomposition} and projectors \eqref{Eq:Projectors} it takes the form
\begin{equation}
\begin{split}
S&=-M_p^3N_c^2\int \frac{d^4q}{(2\pi)^2}\,\bigg(
{h_b}_{\mu\nu}\frac{\Pi^{\mu\nu,\alpha\beta}\mathcal{F^{TT}}(q)}
{(q^2)^2}{h_b}_{\alpha\beta}+\\
&\frac{16}{3}\varphi_b\,\mathcal{F}_s(q) \varphi_b
+\varphi_{b}\left[\frac{P_{T}^{\mu\nu}\mathcal{F}^{T}(q)}{3q^2}+
\frac{P_{L}^{\mu\nu}\mathcal{F}^{L}(q)}{q^2}\right]{h_b}_{\mu\nu}
\bigg),
\end{split}
\end{equation}
\noindent
where we have used \eqref{Eq:PhiNPhib}, \eqref{Eq:hNphib} and \eqref{Eq:hNhb}. The two-point functions are determined through functional derivative of this result, thus, for the scalar operator we get
\noindent
\begin{equation}\label{Eq:TwoPointScalar}
\left\langle\mathcal{O}\mathcal{O}\right\rangle=
\frac{\delta^2 S}{\delta\varphi_b\delta\varphi_b}\propto 
M_p^3N_c^2\mathcal{F}_s(q)=\mathcal{K} M_p^3N_c^2\frac{1}{q^2},
\end{equation}
\noindent
where the constant is given by $\mathcal{K}=-8192\left(3^3\Gamma(5/4)/W_{\infty}^2 -2^{17/4}\Gamma(3/4)\right)^{-1}$. This result shows us a pole at $q^2=0$, thus, confirms the result obtained in \eqref{Eq:TwoPointScalarQN} and the existence of a massless pole in the spectrum, see Fig.~\ref{Fig:ScalarMass}.
\noindent

On the other hand, the two-point function for the energy-momentum tensor is given by
\noindent
\begin{equation}
\left\langle T^{\mu\nu}T^{\alpha\beta}\right\rangle=
\frac{\delta^2 S}{\delta {h_b}_{\mu\nu}\delta{h_b}_{\alpha\beta}}\propto 
M_p^3N_c^2\frac{\Pi^{\mu\nu,\alpha\beta}}{(q^2)^2}\mathcal{F}^{TT}(q),
\end{equation}
\noindent
as $\mathcal{F}^{TT}(q)$ is, in fact, a constant (see result \eqref{Eq:FTT}), there is no pole on the two-point function of this sector, which is in agreement with the spectrum of the spin 2 sector, see Fig.~\ref{Fig:TensorMass}. Finally, we obtain the mixed two-point function
\noindent
\begin{equation}
\left\langle T^{\mu\nu}\mathcal{O}\right\rangle
\propto 
M_p^3N_c^2\left[\frac{P_T^{\mu\nu}}{3q^2}\mathcal{F}^{T}(q)+
\frac{P_L^{\mu\nu}}{q^2}\mathcal{F}^{L}(q)\right],
\end{equation}
\noindent
as the functions $\mathcal{F}^{T}(q)$ and $\mathcal{F}^{L}(q)$ are constants, the two-point function has a pole at $q^2=0$, this pole is expected because we are mixing the spin-zero and two correlation functions \cite{Hoyos:2013gma}.

\section{Conclusion and final remarks}
\label{Sec:Conclusion}

In this work we investigated the conformal symmetry breaking using a simple bottom-up holographic model. We propose to consider the dilaton field as an input to solve the differential equation describing the background. The dilaton field in the UV guarantees the correct coupling between the dilaton and the corresponding dual operator in the dual field theory $\mathcal{O}$. 
On the other hand, in the IR region, the dilaton guarantees color confinement. To recover the Regge behavior, i.e., $m^2\propto n$, the dilaton must be quadratic. Moreover, considering the linear dilaton, the Regge behavior is not guaranteed and the spectrum becomes a continuum. Solving the perturbation equations we found the spectrum of the scalar and tensor sectors as a function of the conformal dimension $\epsilon$ (c.f. Figs.~ \ref{Fig:ScalarMass} and \ref{Fig:TensorMass}). We observed that in the limit of vanishing conformal dimension a massless mode arises in the scalar sector. This state may be interpreted as a Nambu-Goldstone boson arising due to the spontaneous conformal symmetry breaking. We confirm that this massless mode is, in fact, a Nambu-Goldsonte boson because the VEV of the corresponding operator is not zero in this limit, i.e., $\langle\mathcal{O}\rangle\neq0$, while the trace of the energy-momentum tensor vanishes $\langle {T}^{\mu}_{\mu}\rangle=0$. We also showed that the massless mode becomes the lightest state when explicit conformal symmetry breaking happens. Additionally, we found an analytic expression for the mass of the lightest state as a function of the conformal dimension, $m_s^2\sim \epsilon^{4/5}$. Finally, we point out that the relation between the mass of the dilaton field leading to explicit breaking of conformal symmetry, and consequently a massive scalar state in the dual field theory, is true in bottom-up holographic models at zero temperature where the CFT is deformed by a relevant operator, and confinement is guaranteed in the IR region. However, this conclusion is not true when a black hole is embedded in the bulk gravity or even though when charge is included in the five-dimensional action, which is equivalent to add chemical potential in the dual field theory.

As a complementary analysis, in the second part of this work, we compute the two-point correlation functions of the dual operators associated with the scalar and tensor perturbations in the bulk gravity. To do so, we expanded the on-shell action up to second order in the perturbations. Finally, to get the desired functions we derive with respect to the source. For the scalar operator, we observed that the two-point function has a pole at $q^2=0$, which represents the massless mode emerging in the spectrum. On the other hand, the two-point function of the tensor sector does not have any pole, which is in agreement with the spectrum. Additionally, we have found the mixed two-point function $\langle{T}^{\mu\nu}\mathcal{O}\rangle$, which has a pole at $q^2=0$, confirming the existence of a massless state in the limit of $\epsilon\to 0$. Future perspectives may investigate the RG flow of the scalar operator at zero and finite temperature, also the relation between the mass of the dilaton field and conformal symmetry breaking.

\section*{Acknowledgments}

We would like to thank Carlos Hoyos and Alfonso Ballon Bayona for stimulating discussions along the development of this work. We also thank Alex Miranda and Alfonso Ballon Bayona for reading the manuscript and recommendations. The author has financial support from Coordena\c{c}\~ao de Aperfei\c{c}oamento do Pessoal de N\'ivel Superior - Programa Nacional de P\'os-Doutorado (PNPD/CAPES, Brazil).

\appendix

\section{Perturbation equations}\label{AppendixA}

In this Appendix, we write the equations obtained after perturbing the background metric and scalar field. We write the equations in terms of the extrinsic curvature and induced metric, both are represented with $K_{\mu\nu}$ and $g_{\mu\nu}$, respectively, while Greek letters characterize the indices. In terms of the domain wall coordinates the extrinsic curvature is given by
\noindent
\begin{equation}\label{Eq:ExtrinsecCurv}
\begin{split}
K_{\mu\nu}&=-\Gamma^{r}_{\mu\nu}=\frac{1}{2}\partial_{r}g_{\mu\nu},\\
K^{\mu}_{\nu}&=\Gamma^{\mu}_{r\,\nu},\quad K=g^{\mu\nu}\,K_{\mu\nu}.
\end{split}
\end{equation}
\noindent
In terms of the extrinsic curvature the components of the background equations \eqref{Eq:BackgroundEqs} may be written as:
\noindent
\begin{equation}
\begin{aligned}
\label{Eq:FluctuationsV1N1}
-\partial_r K-K^{\alpha}_{\beta}K^{\beta}_{\alpha}
=\frac{4}{3}(\partial_r\Phi)^2-\frac{1}{3}V,\\
\end{aligned}
\end{equation}
\begin{equation}
\begin{aligned}
\label{Eq:FluctuationsV1N2}
&-\partial_{\mu}K+\nabla_{\alpha}K^{\alpha}_{\mu}
=\frac{4}{3}(\partial_r\Phi)\partial_{\mu}\Phi,\\
\end{aligned}
\end{equation}
\begin{equation}
\begin{aligned}
\label{Eq:FluctuationsV1N3}
&R_{\mu\nu}-\partial_r K_{\mu\nu}+2K^{\alpha}_{\mu}K^{\alpha}_{\nu}
-K K_{\mu\nu}=\frac{4}{3}(\partial_{\mu}\Phi)(\partial_{\nu}\Phi)\\
&-\frac{1}{3}g_{\mu\nu}V
+\frac{4}{3}(\partial_{\mu}\Phi)(\partial_{\nu}\Phi),\end{aligned}
\end{equation}
\begin{equation}
\begin{aligned}
\label{Eq:FluctuationsV1N4}
&\partial^2_{r}\Phi+K\partial_r\Phi
+\partial_{\mu}\left(g^{\mu\nu\partial_{\nu}\Phi}\right)
+\Gamma^{\alpha}_{\mu\alpha}\left(g^{\mu\nu}\partial_{\nu}\Phi\right)\\
&+\frac{3}{8}\partial V=0,
\end{aligned}
\end{equation}
\noindent
where $\nabla_{\alpha}$ and $R_{\mu\nu}$ are the covariant derivative and Riemann tensor depending on the induced metric $g_{\mu\nu}$. We point out that the background equations written in terms of the extrinsic curvature may be useful when studying renormalization group flow equations (see for instance Refs.~\cite{Skenderis:2002wp,Papadimitriou:2004rz,deBoer:1999tgo,Bianchi:2001de,Kiritsis:2014kua}). For the forthcoming analysis, it will be useful to eliminate terms containing $\partial_r K$ and $R_{\mu\nu}$ from the equations above, thus, the background equations become
\noindent
\begin{equation}
\begin{aligned}
\label{Eq:FluctuationsV2N1}
K^{\mu}_{\nu}K^{\nu}_{\mu}-K^2+\frac{4}{3}(\partial_r\Phi)^2
=&\frac{4}{3}\left(\partial^{\mu}\Phi\right)\left(\partial_{\mu}\Phi\right)
\\&-V-R,
\end{aligned}
\end{equation}
\begin{equation}
\label{Eq:FluctuationsV2N2}
\begin{aligned}
-\partial_{\mu}K+\nabla_{\alpha}K^{\alpha}_{\mu}
=\frac{4}{3}(\partial_r\Phi)\partial_{\mu}\Phi,
\end{aligned}
\end{equation}
\begin{equation}
\begin{aligned}
\label{Eq:FluctuationsV2N3}
-\partial_r K^{\mu}_{\nu}-K K^{\mu}_{\nu}
=&\frac{4}{3}(\partial^{\mu}\Phi)(\partial_{\nu}\Phi)
\\&-\frac{1}{3}\delta^{\mu}_{\nu}V-R^{\mu}_{\nu},
\end{aligned}
\end{equation}
\begin{equation}
\begin{aligned}
\label{Eq:FluctuationsV2N4}
\partial^2_{r}\Phi+K\partial_r\Phi
+\partial_{\mu}\left(g^{\mu\nu\partial_{\nu}\Phi}\right)
&+\Gamma^{\alpha}_{\mu\alpha}\left(g^{\mu\nu}\partial_{\nu}\Phi\right)\\
&+\frac{3}{8}\partial V=0,
\end{aligned}
\end{equation}
\noindent
Now introducing the perturbations on the background metric and scalar field defined in \eqref{Eq:Perturbations}. Under this definition, the Ricci tensor and its scalar to linear-order in the perturbations take the form 
\noindent
\begin{equation}
\begin{split}
R_{\mu\nu}^{\scriptscriptstyle{(1)}}
&=\frac{1}{2}\left(\partial_{\mu}\partial_{\alpha}h^{\alpha}_{\nu}
+\partial_{\nu}\partial_{\alpha}h^{\alpha}_{\mu}
-\partial_{\nu}\partial_{\mu}h-\square h_{\mu\nu}\right),\\
R^{\scriptscriptstyle{(1)}}&=
e^{-2A}\left(\partial_{\mu}\partial_{\nu}h^{\mu\nu}\right),
\end{split}
\end{equation}
\noindent
where $\square=\partial^{\mu}\partial_{\mu}$. Plugging these results in  \eqref{Eq:FluctuationsV2N1} $-$ \eqref{Eq:FluctuationsV2N4} and applying the Fourier's transform we get the following equations:
\noindent
\begin{equation}
\begin{aligned}
\label{Eq:FluctuationsV3N1}
3\partial_{r}A \,\partial_{r}h+\frac{8}{3}\partial_{r}\Phi_0\partial_{r}\varphi
+\partial V\,\varphi-e^{-2A}h_{T}=0,
\end{aligned}
\end{equation}
\begin{equation}
\label{Eq:FluctuationsV3N2}
\begin{aligned}
q_{\mu}\partial_{r}h-q_{\alpha}\partial_{r}h^{\alpha}_{\mu}
-\frac{8}{3}q_{\mu}\partial_{r}\Phi_{0}\,\varphi=0,
\end{aligned}
\end{equation}
\begin{equation}
\begin{aligned}
\label{Eq:FluctuationsV3N3}
&\partial^2_r h^{\mu}_{\nu}+4\partial_{r}A \partial_{r}h^{\mu}_{\nu}
+\left(\partial_{r}A\partial_{r}h+\frac{2}{3}\partial V\,\varphi\right)\delta^{\mu}_{\nu}\\
&-e^{-2A}\left((q^{\mu}\eta^{\alpha\beta}
-q^{\alpha}\eta^{\mu\beta})q_{\nu}h_{\alpha\beta}+
P_{T}^{\alpha\mu}h_{\alpha\nu}\right)
=0,
\end{aligned}
\end{equation}
\begin{equation}
\begin{aligned}
\label{Eq:FluctuationsV3N4}
&\partial^2_{r}\Phi+4\partial_{r}A\,\partial_{r}\varphi
+\frac{1}{2}\partial_{r}\Phi\partial_{r}h-q^2e^{-2A}\varphi\\
&+\frac{3}{8}\partial^2 V \varphi=0.
\end{aligned}
\end{equation}
\noindent
The next stage is to project some of the last equations along the momentum and transverse to it (see Refs.~\cite{Hoyos:2012xc,Mueck:2001cy}). For doing that we decompose the metric perturbations in the form  
\noindent
\begin{equation}\label{Eq:hDecomposition}
h_{\mu\nu}=h_{\mu\nu}^{TT}+h_{\mu\nu}^{TL}+h_{\mu\nu}^{T}+h_{\mu\nu}^{L}.
\end{equation}
\noindent
The projectors are defined by 
\noindent
\begin{equation}\label{Eq:Projectors}
\begin{split}
P_{T}^{\mu\nu}=&\,q^2\eta^{\mu\nu}-q^{\mu}q^{\nu},\\
P_{L}^{\mu\nu}=&\,q^{\mu}q^{\nu},\\
\Pi^{\mu\nu,\alpha\beta}=&\,P_{T}^{\mu\alpha}P_{T}^{\nu\beta}-
\frac{1}{3}P_{T}^{\mu\nu}P_{T}^{\alpha\beta},
\end{split}
\end{equation}
\noindent
where $P_{L}^{\mu\nu}$ projects along $q^{\mu}$, $P_{T}^{\mu\nu}$ projects in the transverse direction and $\Pi^{\mu\nu,\alpha\beta}$ projects on the transverse-traceless sector, the components of \eqref{Eq:hDecomposition} are given by \noindent
\begin{equation}
\begin{split}
h^{TT}_{\mu\nu}=&\frac{1}{(q^2)^2}\Pi_{\mu\nu}^{\alpha\beta}h_{\alpha\beta},\\
h^{TL}_{\mu\nu}=&\frac{1}{(q^2)^2}\left[{P_T}^{\alpha}_{\mu}{P_L}^{\beta}_{\nu}
+{P_T}^{\beta}_{\nu}{P_L}^{\alpha}_{\mu}\right]h_{\alpha\beta},\\
h^{T}_{\mu\nu}=&\frac{1}{3(q^2)^2}{P_T}_{\mu\nu}{P_T}^{\alpha\beta}h_{\alpha\beta},\\
h^{L}_{\mu\nu}=&
\frac{1}{(q^2)^2}{P_L}^{\alpha}_{\mu}{P_L}^{\beta}_{\nu}h_{\alpha\beta}.
\end{split}
\end{equation}
\noindent
Additional properties of the projectors are: $h_L={P_L}^{\alpha\beta}h_{\alpha\beta}/q^2$, $h_T={P_{T}}^{\alpha\beta}h_{\alpha\beta}/q^2$, $h_L+h_T=h$ and ${P_L}^{\mu\nu}/q^2+{P_T}^{\mu\nu}/q^2=\eta^{\mu\nu}$.

Let us split the problem. First, we apply the projector $P_{L}^{\mu\nu}$ on \eqref{Eq:FluctuationsV3N2} to get 
\noindent
\begin{equation}
\begin{aligned}
\label{Eq:FluctuationsV4N2}
\partial_{r}h_{T}+\frac{8}{3}\partial_{r}\Phi\,\varphi=0.
\end{aligned}
\end{equation}
\noindent
Second, we apply the projectors ${P_{L}}^{\nu}_{\mu}$ and ${P_{T}}^{\nu}_{\mu}$ 
on \eqref{Eq:FluctuationsV3N3}, we get the following equations
\noindent
\begin{equation}\label{Eq:hLhTEqs}
\begin{split}
&\partial_r^2 h_L+4\partial_r A\partial_r h_L+\partial_r A \partial_r h
+\frac23\partial V\,\varphi-q^2e^{-2A}h_T=0,\\
&\partial_r^2 h_T+4\partial_r A\partial_r h_T+3\partial_r A \partial_r h
+2\partial V\,\varphi-q^2e^{-2A}h_T=0.
\end{split}
\end{equation}
\noindent
Combining both equations, then using \eqref{Eq:FluctuationsV3N1} to replace $h_T$, at the end we get a second-order differential equation for $h$ 
\noindent
\begin{equation}
\partial_r^2 h+2\partial_r A\partial_r h
+\frac23\partial V\,\varphi
-\frac{16}{3}\partial_r\Phi_0\partial_r \varphi=0,
\end{equation}
\noindent
which may be written as 
\begin{equation}
e^{-2A}\partial_r\left(e^{2A}\partial_r h\right)
+\frac23\partial V\,\varphi
-\frac{16}{3}\partial_r\Phi_0\partial_r \varphi=0.
\end{equation}
\noindent
Defining the new function $H=e^{2A}\partial_r h$, the last equation becomes a first-order differential equation for $H$. Additionally, we do introduce this new function in the Klein-Gordon equation \eqref{Eq:FluctuationsV3N4}, the resulting equations are:
\begin{equation}
\begin{aligned}
\label{Eq:FluctuationsV4N3}
e^{-2A}\partial_{r}H+\frac{2}{3}\partial V\varphi
-\frac{16}{3}\partial_{r}\Phi\partial_{r}\varphi=0,\\
\end{aligned}
\end{equation}
\begin{equation}
\begin{aligned}
\label{Eq:FluctuationsV4N4}
&\partial^2_{r}\Phi+4\partial_{r}A\,\partial_{r}\varphi
+\frac{1}{2}\partial_{r}\Phi e^{-2A}H
-q^2e^{-2A}\varphi\\&+\frac{3}{8}\partial^2 V \varphi=0.
\end{aligned}
\end{equation}
\noindent
We may eliminate the dependence on $H$ using both equations, the final result is a third-order differential equation \eqref{Eq:ThirdDiffEqScalar}.

On the other hand, to complement the analysis we write the equation of the transverse and traceless sector, which is obtained by applying the projector ${\Pi^{\alpha\beta,\nu}}_{\mu}$ on \eqref{Eq:FluctuationsV3N3}, hence, we get
\noindent
\begin{equation}\label{Eq:hTTEq}
\partial^2_rh^{TT}_{\mu\nu}+4\partial_rA\partial_rh^{TT}_{\mu\nu}
-q^2e^{-2A}\,h^{TT}_{\mu\nu}=0.
\end{equation}
\noindent
This equation is used in Section \ref{Sec:TensorPerturb} to find the two-point function of the energy-momentum tensor.

\section{Scalar perturbations - linear dilaton}
\label{Sec:AppendixB}

In this Appendix, we write the results obtained for the linear dilaton in the IR. Thus, performing some integrals in \eqref{Eq:PertSol} the perturbative solution for $\alpha=1$ is given by 
\noindent
\begin{equation}\label{Eq:PerturbaSolLinear}
\begin{split}
\mathcal{S}=&c_2\left(1+\frac{9\widetilde{q}^{\,2}}{8}\Phi+\mathcal{O}(\widetilde{q}^{\,4})\right)-\\
&\frac{c_1e^{-2\Phi}}{3W^2_{\infty}}\left(1-
\frac{9\widetilde{q}^{\,2}}{8}\Phi+\mathcal{O}(\widetilde{q}^{\,4})\right).
\end{split}
\end{equation}
\noindent

On the other hand, we get an analytic solution of Eq.~\eqref{Eq:ScalarIR}, which is given by 
\noindent
\begin{equation}\label{Eq:IRAsympSol1}
\begin{split}
\mathcal{S}=&c_3\,e^{\left(-2-\sqrt{4+9\widetilde{q}^{\,2}}\right)\frac{\Phi}{2}}+
c_4\,e^{\left(-2+\sqrt{4+9\widetilde{q}^{\,2}}\right)\frac{\Phi}{2}}.
\end{split}
\end{equation}
\noindent
Expanding the last result up to second order in $\widetilde{q}^{\,2}$ 
\noindent
\begin{equation}\label{Eq:AnaSolLinear}
\begin{split}
\mathcal{S}=&c_4\left(1
+\frac{9\widetilde{q}^{\,2}}{8}\Phi+\mathcal{O}(\widetilde{q}^{\,4})\right)+\\
&c_3e^{-2\Phi}\bigg(1-\frac{9\widetilde{q}^{\,2}}{8}\Phi
+\mathcal{O}(\widetilde{q}^{\,4})\bigg).
\end{split}
\end{equation}
\noindent
Matching \eqref{Eq:PerturbaSolLinear} with \eqref{Eq:AnaSolLinear} 
we get 
\noindent
\begin{equation}\label{Eq:c2c1IRRatioLinearDil}
\frac{c_2}{c_1}=-\frac{c_4}{3\,c_3W_{\infty}^2}.
\end{equation}
\noindent
On the other hand, the solution close to the boundary is the same as in \eqref{Eq:UVDilaExp}. Hence, combining Eqs.~\eqref{Eq:c2c1IRRatioLinearDil}, \eqref{Eq:UVDilaCoeff} and \eqref{Eq:CNC1} we get the following relation
\noindent
\begin{equation}\label{Eq:PhiNPhibLinearDil}
\varphi_N=\mathcal{F}_s(q)\,\varphi_b,
\end{equation}
\noindent
where
\noindent
\begin{equation}
\mathcal{F}_s(q)=-\frac{512}{\left(\frac{3^2 c_4}{c_3W_{\infty}^2}
-\frac{2^{17/4}\Gamma(3/4)}{3}\right)}\frac{1}{q^2}.
\end{equation}
\noindent
Using the same idea as we have done for the quadratic dilaton, the solution close to the boundary may be written as in \eqref{Eq:QNMApproach}, from this expression we read the coefficients. Thus, the two-point function for the linear dilaton is given by 
\noindent
\begin{equation}
\langle\mathcal{O}\mathcal{O}\rangle\propto\frac{\varphi_{N}}{\varphi_b}
\propto\frac{1}{q^2}.
\end{equation}
\noindent
Therefore, we have shown that the correlation function has a pole at $q^2=0$, which corresponds to the massless state displayed in Fig.~\ref{Fig:ScalarMass}.

\end{document}